\documentclass[twocolumn]{aastex701}
\usepackage{amsmath}
\newcommand{\demp}{DEmP2026}
\newcommand{\pdz}{PDz}
\newcommand{\snmad}{\sigma_{\rm NMAD}}
\newcommand{\fout}{f_{\rm out}}

\makeatletter\def\raggedcolumn@sw#1#2{#1}\makeatother

\shorttitle{DEmP2026 on HSC PDR3}
\shortauthors{Hsieh}

\begin{document}

\title{DEmP2026: calibrated photometric-redshift, stellar-mass and star-formation-rate distributions from a generative network, applied to the Hyper Suprime-Cam Public Data Release 3}

\author[0000-0001-5615-4904]{Bau-Ching Hsieh}
\affiliation{Academia Sinica Institute of Astronomy and Astrophysics, 11F of AS/NTU Astronomy-Mathematics Building, No.1, Sec. 4, Roosevelt Rd, Taipei 106319, Taiwan}
\email{bchsieh@asiaa.sinica.edu.tw}

\begin{abstract}
I present \demp, an empirical code that estimates the probability distributions of the photometric redshift (\pdz), stellar mass and star-formation rate (SFR) of galaxies from broad-band photometry and morphology. Its core is engression. Neural networks take random numbers as an extra input and turn them into possible redshifts, and they are trained with the energy score, a loss that is smallest when the generated redshifts follow the true distribution. The \pdz\ are then calibrated, and a second set of faster networks, trained also on millions of galaxies without redshifts, reproduces them in a single evaluation. I apply \demp\ to the Deep/UltraDeep and Wide layers of the Hyper Suprime-Cam Subaru Strategic Program Public Data Release 3, trained only on public spectroscopic redshifts and COSMOS-Web photometric redshifts. For secure spectroscopic redshifts, \demp\ reaches $\snmad = 0.014$ (Deep/UltraDeep) and 0.018 (Wide), with outlier fractions of 3.4\% and 3.9\%. For COSMOS-Web galaxies to $i \approx 27$, it reaches 0.079. In both layers, the 68\% and 95\% intervals contain the reference redshift 67\% and 95\% of the time. Trained only on bright or only on blue galaxies, the networks still predict faint or red ones with $\snmad = 0.021$ and 0.033. Compared with the original DEmP photo-$z$ of the same release, \demp\ reduces $\snmad$ by 35\% (Deep/UltraDeep) and 31\% (Wide) and the outlier fraction by 35\% and 52\%. The same code recovers the stellar masses and SFRs of DESI DR1 with scatters of 0.13--0.14 and 0.18--0.19\,dex, and those of COSMOS-Web with 0.26--0.33 and 0.28--0.41\,dex.
\end{abstract}


\section{Introduction} \label{sec:intro}

Photometric redshifts are the only practical way to place the hundreds of millions of galaxies of modern imaging surveys along the line of sight. Photometric-redshift methods fall into two families \citep{Salvato2019, Newman2022}. Template-fitting codes compare the observed fluxes with redshifted model or empirical spectra \citep[e.g.,][]{Benitez2000, Ilbert2006, Brammer2008, Tanaka2015}. Empirical codes learn the relation between photometry and redshift from galaxies with known redshifts, using nearest neighbors, decision trees or neural networks \citep[e.g.,][]{Collister2004, CarrascoKind2013, Sadeh2016, Pasquet2019}. When the training set is representative of the target galaxies, empirical codes are usually more accurate than template fitting. However, the reliability of their photometric-redshift probability distributions (\pdz) depends on how each code constructs them, and it has to be tested.

For most scientific uses, the \pdz\ matters as much as the point estimate, and a \pdz\ is useful only if it is calibrated. This means that its 68\% interval should contain the true redshift for 68\% of the galaxies, at every magnitude and redshift. Calibration is usually tested with the probability integral transform (PIT). Many codes give \pdz\ that are too narrow or too broad \citep{Bordoloi2010, Polsterer2016, Schmidt2020}, which has motivated recalibration of the \pdz\ after they are computed \citep{Bordoloi2010, Dey2021}. Better point estimates can also be extracted from existing \pdz\ \citep{Duarte2026}, a study that included the PDR3 \pdz\ of the original DEmP.

The Direct Empirical Photometric method \citep[DEmP;][]{Hsieh2014} produced the photometric redshifts of several data releases of the Hyper Suprime-Cam Subaru Strategic Program (HSC-SSP) \citep{Tanaka2018, Nishizawa2020}. In this paper, I present its successor, \demp. It keeps the empirical approach of DEmP and the calibration of the \pdz\ against training galaxies that were not used to fit them. However, it replaces the estimator by neural networks that generate samples of the redshift distribution. It also adds faster networks that reproduce the results of the first networks and learn from galaxies without redshifts as well. The same code gives stellar masses and star-formation rates (SFRs).

To make every result reproducible, I use only public data. These are the photometry of the HSC-SSP Public Data Release 3 \citep[PDR3;][]{Aihara2022}, public spectroscopic redshifts, and the public COSMOS-Web multi-band photometric-redshift catalog. Section~\ref{sec:data} describes the data and the training and test sets, and Section~\ref{sec:method} describes the method. Section~\ref{sec:photoz} presents the photometric-redshift results. Section~\ref{sec:od} tests how well the networks predict galaxies unlike those of the training set. There, the networks are trained only on bright galaxies and predict only faint ones, or are trained only on blue galaxies and predict only red ones. Section~\ref{sec:orig} compares the results with the original DEmP photo-$z$ of PDR3. Section~\ref{sec:mass} presents the stellar masses and SFRs, Section~\ref{sec:compute} the computational cost, and Section~\ref{sec:summary} gives a summary.

\section{Data} \label{sec:data}

\subsection{HSC PDR3 photometry} \label{sec:hsc}

The HSC-SSP (\citealt{Aihara2018a}; \citealt{Aihara2018b}; \citealt{Aihara2019}) is observed with Hyper Suprime-Cam \citep{Miyazaki2018, Komiyama2018, Kawanomoto2018, Furusawa2018}. Its data are processed with the HSC pipeline (\citealt{Bosch2018}; \citealt{Bosch2019}), which builds on the LSST software \citep{Juric2017, Ivezic2019}. PDR3 \citep{Aihara2022} provides $grizy$ imaging in three layers. The Deep and UltraDeep layers (DUD hereafter) cover 37\,deg$^2$ to a $5\sigma$ point-source depth of $i \approx 26.9$. The Wide layer covers about 670\,deg$^2$ at full depth in all five bands (1,470\,deg$^2$ including partially observed area), to $i \approx 26.2$.

I use the forced photometry of primary objects from the reprocessed DUD data of the public release (database schema \texttt{pdr3\_dud\_rev}) and from the Wide catalog. The reprocessing added exposures that were missing from the first processing of a few DUD tracts. The magnitudes are corrected for three effects. The first is Galactic extinction, which the release computes from the map of \citet{Schlegel1998}. The second is the zero-point offset derived from the stellar sequence in each patch. The third is the change of the $r$ and $i$ filters during the survey. The last two corrections are published as separate tables of the release, and I subtract them as the release documentation recommends \citep{Aihara2022}.

\subsection{Input quantities} \label{sec:features}

I derive the photometric redshift of each object from 77 input quantities (Table~\ref{tab:features}). These are magnitudes and colors in several types of photometry (i.e., undeblended and deblended PSF-matched convolved apertures, fixed circular apertures, cModel, PSF, Kron, and the exponential and de Vaucouleurs components of cModel). They also include morphological quantities, namely the sizes and shapes from the second moments corrected for the PSF, the concentration, the fraction of the de Vaucouleurs component, the ellipticity, the PSF size in each band, the blendedness, the Kron radius and the footprint area. For 16 of these quantities, the measurement errors are used in two ways (Table~\ref{tab:features}). First, the logarithms of these errors are additional inputs of the networks. Second, the errors set the size of the perturbations applied to the inputs during training (Section~\ref{sec:engression}). A measurement that is missing or not finite is marked as missing, and the networks learn to handle missing values (Section~\ref{sec:missing}).

For the stellar masses and SFRs, I use 83 input quantities. In addition to the 77 quantities above, they include, in each of the five bands, the difference between the undeblended and deblended 1.5\arcsec\ convolved magnitudes. This difference measures the light that neighboring objects add to the aperture. The $i$-band difference is also placed among the main input quantities, which are used to measure the distance to the training set (Section~\ref{sec:ood}). It therefore enters twice, which gives 83 quantities in total.

\begin{deluxetable}{lll}
\tablecaption{Input quantities for the photometric redshifts. \label{tab:features}}
\tablehead{\colhead{Columns} & \colhead{Quantity} & \colhead{Error}}
\startdata
0--4   & undeblended convolved 1.5\arcsec\ $grizy$ magnitudes & yes \\
5--8   & their colors & yes \\
9      & $i$ PSF-corrected size & yes \\
10     & $i$ cModel magnitude & yes \\
11     & concentration & yes \\
12     & $i$ fracDev & no \\
13     & $i$ ellipticity & no \\
14--17 & deblended convolved 1.5\arcsec\ colors & yes \\
18--21 & deblended convolved 1.1\arcsec\ colors & no \\
22--25 & cModel colors & no \\
26--29 & PSF colors & no \\
30--33 & $g, r, z, y$ sizes & no \\
34--38 & Kron $-$ convolved magnitude & no \\
39--43 & cModel $-$ convolved magnitude & no \\
44--48 & log PSF size & no \\
49--56 & exponential and de Vaucouleurs colors & no \\
57--60 & Kron colors & no \\
61--64 & undeblended convolved 2.0\arcsec\ colors & no \\
65--69 & undeblended aperture 1.0\arcsec $-$ 3.0\arcsec\ magnitude & no \\
70--74 & blendedness & no \\
75     & log $i$ Kron radius & no \\
76     & log $i$ footprint area & no \\
\enddata
\tablecomments{Convolved: photometry on images matched to a PSF of 1.34\arcsec\ FWHM, in apertures of the given diameter (database columns \texttt{convolvedflux\_3\_11}, \texttt{\_3\_15} and \texttt{\_3\_20}).}
\end{deluxetable}

\subsection{Reference redshifts} \label{sec:labels}

The spectroscopic redshifts are taken from DESI DR1 \citep{DESI2025}, SDSS DR17 \citep{SDSS_DR17}, GAMA DR4 \citep{Driver2022}, the COSMOS spectroscopic compilation \citep{Khostovan2025}, WiggleZ \citep{Drinkwater2018}, VVDS \citep{LeFevre2013}, OzDES \citep{Lidman2020}, the final release of 6dFGS (\citealt{Jones2004}; \citealt{Jones2009}), UDSz \citep{McLure2013, Bradshaw2013}, VANDELS \citep{Garilli2021}, VUDS \citep{LeFevre2015}, and the LIneA compilation \citep{Gschwend2018}. From each survey, I keep only the redshifts that the survey itself classifies as secure. For DESI, this means \texttt{ZWARN}~$=0$, $\Delta\chi^2>25$ and \texttt{COADD\_FIBERSTATUS}~$=0$. I also use low-precision spectroscopic redshifts from PRIMUS \citep{Coil2011, Cool2013} and from the prism and grism programs in the COSMOS compilation.

I do not use the 2dFGRS redshifts. The public 2dFGRS positions are offset by about $0.7''$ from the HSC positions, and many of them are matched to a fainter neighboring HSC object instead of the 2dFGRS galaxy. Little is lost by this choice, because most bright galaxies in the HSC footprint have a redshift from another survey as well. I also drop the 6dFGS redshifts of objects fainter than $i = 18.5$. The 6dFGS targets are brighter than about $b_J = 16.75$ and $K = 12.65$ \citep{Jones2004}, so an HSC object much fainter than this limit is most likely a wrong match, for example a fainter neighbor or a fragment of a large bright galaxy that the deblender has split.

Each catalog entry is matched to the nearest primary HSC object within 1\arcsec. When several surveys give a redshift for the same object, I keep the redshift if they all agree within $0.005(1+z)$, or if at least two thirds of the surveys (and at least two) agree. Otherwise, the object gets no spectroscopic redshift.

In the COSMOS field, I also use the photometric redshifts of the COSMOS-Web catalog \citep{Casey2023, Shuntov2025}. This catalog combines the JWST imaging of COSMOS-Web with the deep multi-band imaging of the COSMOS field, from the ultraviolet to the mid-infrared. Its photometric redshifts are therefore much more precise than those that five HSC bands can give, and they extend the reference sample to the depth of the HSC photometry. I use them only for objects without a spectroscopic redshift. The COSMOS-Web sources are matched to HSC objects within 0.6\arcsec. When several COSMOS-Web sources fall on one HSC object, I take the redshift of the source that dominates the HSC $i$-band flux. If no source dominates, the object gets no COSMOS-Web redshift. COSMOS-Web sources that are point-like and classified as stars in the catalog are used as stars.

I split the objects with a reference redshift into four classes. Class 1 contains galaxies and quasars with secure spectroscopic redshifts. Class 2 contains galaxies and quasars with low-precision spectroscopic redshifts. Class 3 contains objects with COSMOS-Web photometric redshifts. Class 4 contains the stars, both the spectroscopically confirmed stars and the COSMOS-Web point sources classified as stars. Classes 1 to 3 are used to train and test the photometric redshifts, and class 4 is used for the star classifier (Section~\ref{sec:outputs}).

Table~\ref{tab:samples} lists the number of objects in each class, for the training and test sets of both layers (Section~\ref{sec:design}). In DUD, the COSMOS-Web redshifts (median $i=26.3$) extend the reference sample to the depth of the photometry. In Wide, 3.8 million objects have a class-1 redshift (median $i=21.3$), most of them from DESI DR1.

\begin{deluxetable*}{lrrrrrr}
\tablecaption{Number of objects with reference values in the training and test sets. \label{tab:samples}}
\tablehead{\colhead{} & \multicolumn{3}{c}{DUD} & \multicolumn{3}{c}{Wide} \\ \colhead{} & \colhead{all} & \colhead{training} & \colhead{test} & \colhead{all} & \colhead{training} & \colhead{test}}
\startdata
secure spec-$z$ (class 1)            & 282,636 & 254,419 & 28,217 & 3,823,787 & 3,441,171 & 382,616 \\
low-precision $z$ (class 2)          & 46,599  & 41,967  & 4,632  & 55,371    & 49,852    & 5,519   \\
COSMOS-Web photo-$z$ (class 3)       & 316,736 & 284,958 & 31,778 & 166,057   & 149,280   & 16,777  \\
galaxies + quasars (classes 1--3)    & 645,971 & 581,344 & 64,627 & 4,045,215 & 3,640,303 & 404,912 \\
stars (class 4)                      & 39,081  & 35,220  & 3,861  & 869,092   & 782,188   & 86,904  \\
without redshift (random)            & 3,112,539 & & & 3,739,485 & & \\
\hline
$M_*$, SFR: DESI                     & 123,356 & 110,945 & 12,411 & 2,644,070 & 2,379,612 & 264,458 \\
$M_*$, SFR: COSMOS-Web               & 299,588 & 269,539 & 30,049 & 165,258   & 148,594   & 16,664  \\
\enddata
\tablecomments{The test set is a random 10\% of all objects with a reference value in each layer. Galaxies and quasars include 1,013 (DUD) and 31,458 (Wide) test quasars. Without redshift: the random sample of objects without a reference redshift (Section~\ref{sec:design}). Each reference stellar mass comes with an SFR from the same fit, so the two quantities share one sample for each reference catalog.}
\end{deluxetable*}

\subsection{Reference stellar masses and star-formation rates} \label{sec:masslabels}

For the stellar masses and SFRs, I use two sets of reference values, one from the DESI DR1 physical-properties value-added catalog \citep{Siudek2024} and one from the COSMOS-Web catalog \citep{Shuntov2025}. I keep them apart because they rest on different data and assumptions. The DESI values are CIGALE \citep{Boquien2019} fits. I use them for galaxies whose reference redshift includes a DESI measurement and whose fit redshift agrees with the reference redshift within $0.005(1+z)$. The DESI SFRs are averaged over 10\,Myr. The COSMOS-Web values are the CIGALE stellar masses and 100-Myr SFRs of the catalog. When several COSMOS-Web sources at the reference redshift fall on one HSC object, their masses and SFRs are summed. I drop an object if a COSMOS-Web source within 1\arcsec\ that contributes at least 20\% of its HSC $i$-band flux is a star or lies at a redshift that differs by more than $0.05(1+z)$. The HSC photometry of such an object mixes the light of unrelated sources, so the summed mass and SFR would not describe it.

For galaxies with little star formation, the CIGALE fits can return SFRs many orders of magnitude below any value that broad-band photometry can measure. These values are set by the model grid rather than by the data. If they were kept, the networks would be asked to reproduce numbers that the photometry cannot constrain, and these galaxies would dominate the scatter. For both sets, I therefore floor the SFRs at $\log{\rm SFR} = \log M_* - 12$, which corresponds to a specific SFR of $10^{-12}$\,yr$^{-1}$, and assign an error of 0.5\,dex at the floor. The floor marks these galaxies as quiescent without assigning them a precise SFR.

\subsection{Training set, test set and sample without redshifts} \label{sec:design}

I randomly divide the reference sample into a test set (10\%) and a training set (90\%). The split uses a fixed random seed, so that it can be reproduced. The photometric redshifts, stellar masses and SFRs share this split. A test object is never used to train any of them. All settings of the method were fixed before the test set was evaluated for the first time, so no decision was tuned to the test set. The main results of this paper, in Sections~\ref{sec:photoz}, \ref{sec:orig} and \ref{sec:mass}, come from models trained on the whole training set and evaluated on the test set, in both layers.

In both layers, 5\% of the training galaxies are held back as a validation set during the training of each network (Section~\ref{sec:engression}), and the final results are measured on the test set. In DUD, I also run a ten-fold cross-validation to measure how much the results change with the training set. For this purpose, the training set is divided at random into ten folds of equal size. Each fold is predicted by a model trained on the other nine folds, and the differences between the folds show how sensitive the results are to the training set (Sections~\ref{sec:kfold} and \ref{sec:mass}). In Wide, the results are validated with the test set alone. A single Wide photo-$z$ run takes about eight hours (Section~\ref{sec:compute}), and the Wide test set, with 404,912 galaxies and quasars, is six times larger than the DUD test set.

The sample without redshifts is a random subset of the primary objects that have no reference redshift. It contains about 20\% of the DUD objects with $i<27$ and 1\% of the Wide objects with $i<26$, that is, 3.1 and 3.7 million objects. It is used only to train the fast networks of the photometric redshifts, so that they also see galaxies that are typical of the survey rather than of the spectroscopic samples (Section~\ref{sec:students}).

\section{Method} \label{sec:method}

\subsection{Overview} \label{sec:overview}

\begin{figure*}
\centering
\includegraphics[width=\textwidth]{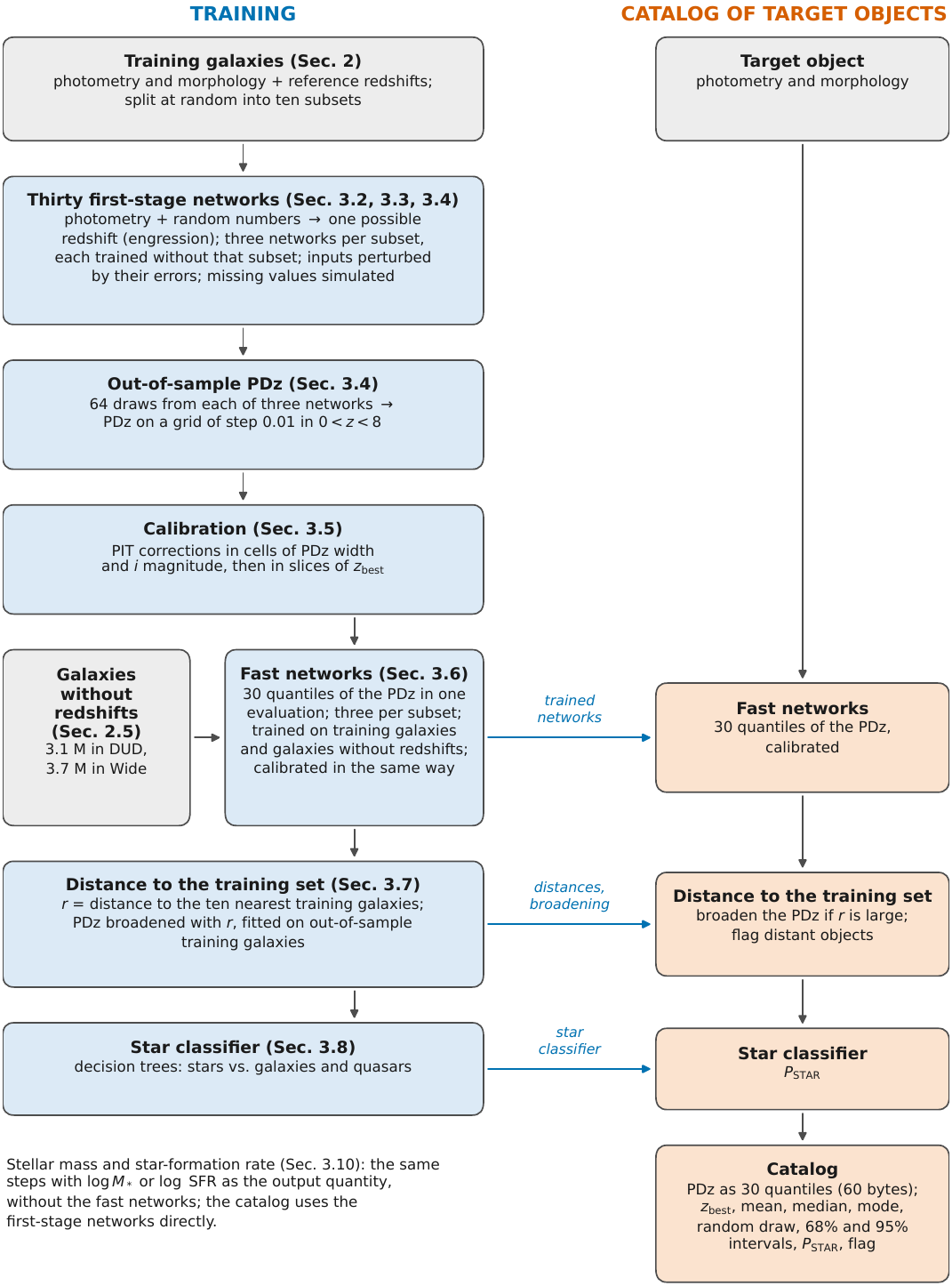}
\caption{The procedure of \demp. Left: training, with the section that describes each step. Right: the production of the catalog for target objects, which uses the networks, distances and classifier obtained in training (blue arrows).} \label{fig:flow}
\end{figure*}

\demp\ works in two stages. In the first stage, a set of neural networks learns the probability distribution of the redshift of a galaxy given its photometry. Each network takes random numbers as an extra input and turns them into one possible redshift. Evaluating the network many times with different random numbers therefore gives a sample drawn from the \pdz. These \pdz\ are then calibrated on the training galaxies. For this purpose, each training galaxy is predicted only by networks that were trained without it. In the second stage, a set of faster networks learns to reproduce the calibrated \pdz\ in a single evaluation, in the form of 30 quantiles. The fast networks are trained on the training galaxies and on millions of galaxies without a redshift. The objects for which the catalog is produced are called the target objects. In this paper, they are the test set. To estimate the photometric redshifts of the target objects, only the fast networks are evaluated, and the first-stage networks are needed only during training. The stellar mass and the SFR are obtained with the same procedure and a different output quantity. Figure~\ref{fig:flow} summarizes the procedure and gives the section in which each step is described.

\subsection{Networks that generate redshifts} \label{sec:engression}

A neural network is a flexible function whose many parameters are fitted to a training set. The first stage of \demp\ uses engression \citep{Shen2024}, a method that lets such a network describe a full probability distribution. The network $g$ receives two inputs. The first is the set of input quantities $\mathbf{x}$ of a galaxy, and the second is a set of random numbers $\boldsymbol{\epsilon}$. The network returns one value $g(\mathbf{x}, \boldsymbol{\epsilon})$ of the quantity to be predicted. For the photometric redshifts, this quantity is $\ln(1+z)$. Photometric-redshift errors grow roughly in proportion to $1+z$, so they are roughly constant in $\ln(1+z)$. The output quantity and all input quantities are shifted and scaled to zero mean and unit variance over the training set. This standard step keeps the numbers inside the network of order unity, which makes the fit stable, and it lets the same settings work for redshift, stellar mass and SFR. The values generated by the network are converted back to redshifts by reversing the shift and the scaling, before the \pdz\ is built from them (Section~\ref{sec:ensemble}).

Different random numbers give different values of $g$ for the same galaxy. The network is trained to make these values follow the distribution of the redshifts of galaxies with the same photometry. In other words, the values generated for a galaxy are a sample drawn from its \pdz.

The quantity minimized in the fit is the energy score. For each training galaxy with true value $y$, the network generates two values, $\hat y_1$ and $\hat y_2$, with independent random numbers. The energy score is
\begin{equation}
L = \tfrac{1}{2}\left(|\hat y_1 - y| + |\hat y_2 - y|\right) - \tfrac{1}{2}\,|\hat y_1 - \hat y_2|,
\end{equation}
averaged over the training galaxies. The first term pulls the generated values toward the true value. The second term pushes them apart, so that they do not all collapse onto a single value. Averaged over many galaxies, $L$ is smallest when the generated values follow the true \pdz\ \citep{Gneiting2007}. The network therefore learns the \pdz\ without any assumption about its shape.

The quantity minimized in the fit, often called the loss function, determines what a network learns. Networks that predict a single value are usually fitted by minimizing the mean squared difference between the predicted and the true values. Such a network predicts approximately the mean redshift of the galaxies with the same photometry, and it gives no \pdz. Networks that predict a \pdz\ are usually fitted by maximizing the probability that the predicted distribution assigns to the true redshift. Some of these codes let the network predict the mean and width of a Gaussian, or of a sum of a few Gaussians \citep[e.g.,][]{DIsanto2018}. The shape of the \pdz\ and the number of its peaks must then be chosen in advance. Other codes divide the redshift range into narrow bins and let the network predict the probability that the redshift falls in each bin, which gives the \pdz\ as a histogram \citep[e.g.,][]{Pasquet2019}. The \pdz\ then cannot be finer than the bins.

Engression with the energy score avoids these limitations. First, the generated values can follow any shape, including several separate peaks, and the \pdz\ can be built on any redshift grid. Second, the energy score depends on the distance between the generated and the true redshifts, so a near miss is penalized much less than a large error. The mean squared difference has the same property, but it describes only a single value and gives no \pdz. The energy score combines this property with a full distribution. Third, the fit aims at the whole distribution rather than at a single best value. As a result, the \pdz\ need only small corrections in the calibration step (Section~\ref{sec:calibration}). In addition, \citet{Shen2024} show that engression predicts beyond the range covered by the training data more reliably than a fit of a single best value, provided that the random scatter enters the relation before a nonlinear transformation. This matters at the faint end, where training galaxies become scarce, and I test it in Section~\ref{sec:od}. Finally, the networks are of the simplest kind. The training time therefore grows only linearly with the size of the training set (Section~\ref{sec:compute}), and the stellar mass and SFR are obtained in the same way by changing only the output quantity.

Each network has seven hidden layers of 1024 units, followed by an output layer that returns the single value $g$. Each hidden layer receives the outputs of the previous layer, or the input quantities in the case of the first layer, together with its own set of 100 random numbers drawn from a standard normal distribution, as in the engression software of \citet{Shen2024}. Every unit of a layer receives all of these values, that is, the 1024 outputs of the previous layer and the 100 random numbers. Before the fit, 5\% of the training galaxies are set aside as a validation set, which is not used to fit the parameters. All other training galaxies are used in the fit. The parameters are fitted in many small steps, as is standard for neural networks. In each pass through the training set, the galaxies are shuffled and divided into batches of 4,096. Each batch gives one step, in which the parameters are changed to lower the energy score of the galaxies in that batch. Every training galaxy is thus used once in every pass. A step on a batch is much faster than a step on the whole training set, and many small steps reach a good fit in far less time. New random numbers are drawn for every galaxy at every step. A network is fitted for at most 300 passes through the training set. The energy score is measured on the validation set during the fit, and the version of the network with the lowest score is kept.

During training, each input that has a measurement error is perturbed by a random amount drawn from that error. The five undeblended 1.5\arcsec\ magnitudes are perturbed by random offsets drawn from their errors, and the four colors formed from them are recomputed from the perturbed magnitudes. The other inputs with errors (i.e., the $i$-band size, cModel magnitude and concentration, and the deblended colors) are perturbed independently of one another. Inputs without errors and missing inputs are not perturbed. The perturbations show the networks how much the input quantities of a galaxy can scatter, so that the \pdz\ of a galaxy with noisy photometry becomes broader.

\subsection{Missing measurements} \label{sec:missing}

About 25\% of the DUD training galaxies and 30\% of the Wide training galaxies (1.09 million of 3.64 million) lack at least one measurement. A missing value is replaced by the median of the training galaxies. In addition, the network receives an indicator for each quantity that is missing in some training galaxies. The indicator is 1 if the value of that quantity is missing and 0 otherwise. The indicators are ordinary inputs of the network, given in addition to the input quantities, and their weights are fitted like all others. Because there is one indicator for each such quantity, the network knows which quantities of a galaxy are missing and can learn how the absence of each one changes the \pdz. The missing value is not estimated from the other quantities.

Some combinations of missing quantities are rare in the training set. To teach the networks to handle every combination, I remove measurements artificially during the fit. In each fitting step, 20\% of the galaxies are given the combination of missing quantities of a randomly chosen incomplete training galaxy. The combinations are drawn with a probability proportional to the square root of their frequency in the training set. This is a compromise between two extremes. Drawing the combinations in proportion to their frequency would rarely show the networks the rare ones. Drawing all combinations equally often would let the thousands of rare combinations dominate the fit at the expense of the common ones.

\subsection{Thirty networks and out-of-sample predictions} \label{sec:ensemble}

The training set is split at random into ten subsets. For each subset, three networks are trained on the other nine subsets, which gives 30 networks in total. Each training galaxy therefore gets predictions from three networks that never saw it during training. These predictions are called out-of-sample predictions in the following. Every other object is assigned at random to one of the ten subsets. The galaxies without redshifts and the target objects of the stellar masses and SFRs are predicted by the three first-stage networks of that subset. The photometric redshifts of the target objects are predicted by the three fast networks of that subset instead (Section~\ref{sec:students}). For each galaxy predicted by the first-stage networks, I draw 64 values from each of its three networks. The 192 values are converted back to redshifts and smoothed into a \pdz\ on a grid of step 0.01 between $z = 0$ and 8.

The three networks of a subset are trained on the same galaxies, but they are not identical. Each network starts from different random initial parameters, sees the training galaxies in a different random order, and receives different random perturbations of its inputs. Their predictions therefore differ slightly, and the differences reflect the uncertainty of the fit itself. Drawing all 192 values from one network would sample only the \pdz\ of that network, including its particular errors. Combining three networks averages over these errors and makes the \pdz\ more robust.

\subsection{Calibration} \label{sec:calibration}

The PIT of a galaxy is the fraction of its \pdz\ that lies below its true redshift. If the \pdz\ of a sample are calibrated, the PIT values of the sample are distributed uniformly between 0 and 1. If the \pdz\ are too narrow, too many PIT values lie near 0 and 1. If they are too broad, too many PIT values lie near 0.5.

The calibration uses the out-of-sample \pdz\ of the training galaxies (Section~\ref{sec:ensemble}). I divide these galaxies into cells, with eight bins of \pdz\ width and eight bins of $i$ magnitude. In each cell, I measure the cumulative distribution of the PIT values. I then transform the cumulative distribution of each \pdz\ in the cell with this measured function. After the transformation, the PIT values of the training galaxies in the cell are uniform. The same transformation is applied to every other galaxy that falls in the cell.

After this first correction, some miscalibration remains within individual photo-$z$ slices. A second correction of the same kind removes it. This correction is measured in eleven slices of $z_{\rm best}$ (Section~\ref{sec:outputs}), again in bins of \pdz\ width. The fast networks (Section~\ref{sec:students}) use only the first correction.

\subsection{Fast networks and galaxies without redshifts} \label{sec:students}

The first-stage networks perform 192 evaluations per galaxy. A second set of networks, called the fast networks, predicts the \pdz\ in a single evaluation. They have the same size as the first-stage networks, and there are again three per subset. Each fast network predicts the calibrated first-stage \pdz\ in the format of the catalog, which consists of 30 quantiles (Section~\ref{sec:q30}).

The fast networks are trained on two samples. The first is the training set, with the calibrated out-of-sample \pdz\ of its galaxies as the target. The second is the sample without redshifts (Section~\ref{sec:design}), with the calibrated \pdz\ that the first-stage networks give for these galaxies as the target. With this second sample, the fast networks see several million galaxies like those to which they will be applied. These galaxies lie mostly at the common redshifts of the survey. To keep the fast networks accurate at rare redshifts, the galaxies are weighted in the fit. For this purpose, I take the median redshift of the first-stage \pdz\ of every galaxy of the two samples and count these medians in bins of 0.1 in $\ln(1+z)$. Each galaxy is then weighted by the inverse of the number of galaxies in its bin, so that galaxies in sparsely populated bins get larger weights. The weights are capped at ten times the median weight, so that a few galaxies at very rare redshifts cannot dominate the fit. After the training, the fast networks are calibrated in the same way as the first stage (Section~\ref{sec:calibration}), using their own out-of-sample predictions for the training galaxies. Training a fast network to reproduce a slower one is known in machine learning as knowledge distillation \citep{Hinton2015}.

The fast networks have three advantages. First, they need one evaluation instead of 192, which makes the production of the catalog fast (Section~\ref{sec:compute}). Second, they give the \pdz\ directly as 30 quantiles, the format of the catalog. Third, through the galaxies without redshifts, they learn the \pdz\ over the whole range of photometry of the survey, not only over that of the training set. In tests during the development of the code, their \pdz\ were as accurate as those of the first-stage networks.

\subsection{Galaxies unlike the training set} \label{sec:ood}

A network cannot tell by itself when a galaxy is unlike all of its training galaxies, and it may then return a \pdz\ that is too narrow. For each galaxy, I therefore measure a distance $r$ to the training set. I use the main input quantities, which are the first 18 quantities in DUD and the first 14 in Wide. Each of them is first scaled to unit variance over the training set, as for the network inputs (Section~\ref{sec:engression}). Without this step, quantities with a large spread would dominate the distance over quantities with a small spread. I then compute the median distance from the galaxy to its ten nearest training galaxies in these quantities. For a training galaxy, only the training galaxies of the other subsets are used. Finally, I divide this distance by its median over all training galaxies. A galaxy with $r = 1$ is thus as close to its training neighbors as the average training galaxy is to its own, and a galaxy with $r = 3$ is three times farther from them.

The \pdz\ are then recalibrated in bins of $r$, with corrections derived from the out-of-sample \pdz\ of the training galaxies as in Section~\ref{sec:calibration}. The last bin starts at the 99.9th percentile of $r$ of the training galaxies, at $r = 12.3$ in DUD and 9.7 in Wide. Beyond it, too few training galaxies are left to measure a correction, so the \pdz\ follows an assumed law. It is stretched about its median by a factor $(r/r_0)^\gamma$ and mixed with the redshift distribution of the training set with a weight that grows as $\beta \ln r$, up to at most 0.5. The value $r_0$ is a high percentile of $r$ of the training galaxies. I chose this form so that the \pdz\ of a more distant galaxy becomes broader and closer to the redshift distribution of the training set, with a slow growth for very distant galaxies. The parameters $r_0$, $\gamma$ and $\beta$ are chosen to minimize the CRPS of the out-of-sample training galaxies with large $r$. In both layers, the fit gives no stretch ($\gamma = 0$) and $\beta = 0.05$, so this last step changes the \pdz\ only slightly. The point estimates are not changed. An object is flagged if its distance, measured in the quantities that it actually has, exceeds that of 99\% of the training galaxies.

\subsection{Point estimates, intervals and stars} \label{sec:outputs}

From each \pdz, I derive the mean, the median, the mode, a random draw, and the 68\% and 95\% intervals, as in the photo-$z$ catalogs of earlier HSC releases \citep{Tanaka2018, Nishizawa2020}. I also derive the best estimate $z_{\rm best}$ of \citet{Tanaka2018}, which minimizes the expected penalty
\begin{equation}
\int p(z)\,\left\{1 - \left[1 + \left(\frac{z - z_{\rm best}}{0.15(1+z)}\right)^2\right]^{-1}\right\}\,dz .
\end{equation}
The penalty levels off for large errors, so the parts of the \pdz\ far from $z_{\rm best}$ carry limited weight.

A separate classifier gives the probability $P_{\rm STAR}$ that an object is a star. It is a gradient-boosted ensemble of decision trees \citep{Friedman2001}, in the histogram-based implementation of scikit-learn \citep{Pedregosa2011}, which follows \citet{Ke2017}. It uses all input quantities and is trained to separate the stars (class 4) from the galaxies and quasars. Its output is calibrated by isotonic regression \citep{Zadrozny2002}, so that, among the objects with a given $P_{\rm STAR}$, that fraction are stars.

\subsection{Probability distributions of redshifts} \label{sec:q30}

The catalog stores each \pdz\ as 30 quantiles. A quantile is the redshift below which a given fraction of the probability lies. For example, the quantile for the fraction 0.5 is the median. The 30 fractions are not evenly spaced. For a Gaussian distribution, they are the fractions that lie below 30 evenly spaced points between $-3$ and $+3$ standard deviations, from 0.13\% to 99.87\%. As a result, the fractions are closer together in the tails than in the core, and the quantiles describe the tails of the \pdz\ as well as its core.

Each quantile is stored as a 16-bit integer, so a \pdz\ takes 60 bytes. The same \pdz\ on the grid of 801 bins, stored as 32-bit numbers, would take 3,204 bytes, so the format is 98\% smaller. It loses little information. The cumulative distributions of a \pdz\ on the grid and of its 30 quantiles differ by at most 0.004 for the median object and by at most 0.02 for 99\% of the objects. The mean CRPS (Section~\ref{sec:photoz}) changes by 0.01\%, and the mean redshift by 0.0004 for the median object.

\subsection{Stellar mass and star-formation rate} \label{sec:massmethod}

For $\log M_*$ and $\log{\rm SFR}$, I use the same first-stage networks, subsets and calibration, with the reference stellar mass or SFR as the output quantity. The distributions are computed on grids from 5 to 13\,dex and from $-7$ to 5.5\,dex, in steps of 0.01\,dex, for $\log M_*$ and $\log{\rm SFR}$, respectively. The best estimate minimizes the same expected penalty as $z_{\rm best}$ (Equation~2), with the width $0.15(1+z)$ replaced by a constant 0.3\,dex. The fast networks and the sample without redshifts are not used for these quantities, so the catalog values come directly from the first-stage networks.

\subsection{Implementation} \label{sec:impl}

\demp\ is written in Python with PyTorch. Every step runs on CPUs or GPUs with the same results to floating-point precision. For 20,000 Wide test objects, the CPU and GPU catalogs agree to $4\times10^{-6}$ in every quantile and give identical $z_{\rm best}$. For the photometric redshifts, an interface to the LSST Redshift Assessment Infrastructure Layers \citep[RAIL;][]{RAIL} gives the same answers as the standalone code, bit for bit.

\section{Photometric redshifts} \label{sec:photoz}

I use the following metrics. The redshift error of a galaxy is $\Delta z = (z_{\rm best} - z)/(1+z)$, where $z$ is the reference redshift. The bias is the median of $\Delta z$, and the scatter is $\snmad = 1.4826\,{\rm median}|\Delta z - {\rm median}\,\Delta z|$. The outlier fraction $\fout$ is the fraction of galaxies with $|\Delta z| > 0.15$. The coverages $C_{68}$ and $C_{95}$ are the fractions of reference redshifts that lie inside the 68\% and 95\% intervals of the \pdz. The continuous ranked probability score \citep[CRPS;][]{Gneiting2007} is the integral of the squared difference between the cumulative \pdz\ and a step function at the reference redshift. It is zero for a perfect \pdz\ and grows with both the offset and the width of the \pdz. Finally, I use the distribution of the PIT values (Section~\ref{sec:calibration}; \citealt{Dawid1984}).

\subsection{Accuracy} \label{sec:accuracy}
\begin{deluxetable*}{llrrrrrrrr}
\tablecaption{Photometric redshifts of the test set. \label{tab:photoz}}
\tablehead{\colhead{Layer} & \colhead{Sample} & \colhead{$N$} & \colhead{Bias} & \colhead{$\snmad$} & \colhead{$\fout$} & \colhead{$C_{68}$} & \colhead{$C_{95}$} & \colhead{CRPS} & \colhead{PIT KS}}
\startdata
DUD  & all                    &  64,627 & $-0.0005$ & 0.0300 & 0.152 & 0.673 & 0.947 & 0.179 & 0.004 \\
     & class 1                &  28,217 & $-0.0012$ & 0.0144 & 0.034 & 0.685 & 0.952 & 0.044 & 0.024 \\
     & class 2                &   4,632 & $+0.0017$ & 0.0226 & 0.069 & 0.595 & 0.905 & 0.054 & 0.080 \\
     & class 3                &  31,778 & $+0.0012$ & 0.0791 & 0.268 & 0.673 & 0.949 & 0.318 & 0.013 \\
     & quasars                &   1,013 & $-0.0075$ & 0.0557 & 0.209 & 0.621 & 0.886 & 0.260 & 0.097 \\
Wide & all                    & 404,912 & $-0.0004$ & 0.0192 & 0.053 & 0.674 & 0.948 & 0.062 & 0.002 \\
     & class 1                & 382,616 & $-0.0004$ & 0.0182 & 0.039 & 0.675 & 0.948 & 0.047 & 0.002 \\
     & class 2                &   5,519 & $+0.0032$ & 0.0258 & 0.068 & 0.601 & 0.887 & 0.060 & 0.107 \\
     & class 3                &  16,777 & $-0.0031$ & 0.1331 & 0.365 & 0.676 & 0.954 & 0.396 & 0.008 \\
     & quasars                &  31,458 & $-0.0053$ & 0.0515 & 0.170 & 0.663 & 0.937 & 0.187 & 0.041 \\
\enddata
\tablecomments{Bias, $\snmad$ and $\fout$ use $z_{\rm best}$. $C_{68}$, $C_{95}$: fraction of reference redshifts inside the central 68\% and 95\% intervals of the \pdz. PIT KS: Kolmogorov--Smirnov distance of the PIT distribution from uniform. Class 1: secure spectroscopic redshifts. Class 2: low-precision redshifts. Class 3: COSMOS-Web photometric redshifts. Classes 1 to 3 include the quasars of that class.}
\end{deluxetable*}

\begin{figure*}
\centering
\includegraphics[width=0.8\textwidth]{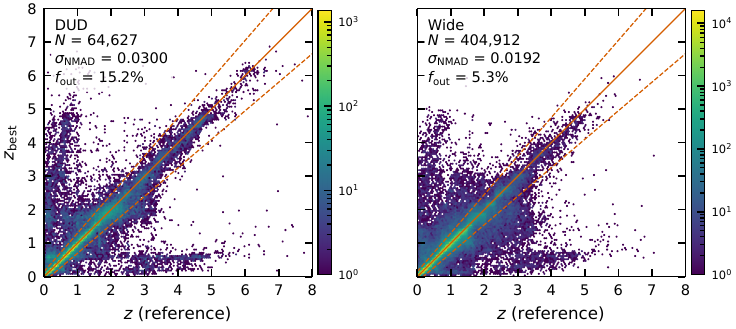}
\caption{$z_{\rm best}$ against the reference redshift for all test galaxies and quasars in DUD (left) and Wide (right). Color: number of objects per hexagonal cell, on a logarithmic scale (empty cells are white). Solid line: $z_{\rm best} = z$. Dashed lines: $|\Delta z| = 0.15$. For display, $z_{\rm best}$, which lies on a grid of step 0.01, is spread uniformly within its grid step.} \label{fig:zz}
\end{figure*}

Table~\ref{tab:photoz} gives the results on the test set, and Figure~\ref{fig:zz} shows $z_{\rm best}$ against the reference redshift. For the secure spectroscopic redshifts (class 1), \demp\ reaches $\snmad = 0.014$ in DUD and 0.018 in Wide. The outlier fractions are 3.4\% and 3.9\%, and the bias is below 0.0015 in absolute value. The COSMOS-Web galaxies (class 3) are much fainter, and half of them in DUD are fainter than $i = 26$. For these galaxies, $\snmad$ is 0.079 in DUD and 0.133 in Wide, where the photometry is shallower.

\begin{figure*}
\centering
\includegraphics[width=\textwidth]{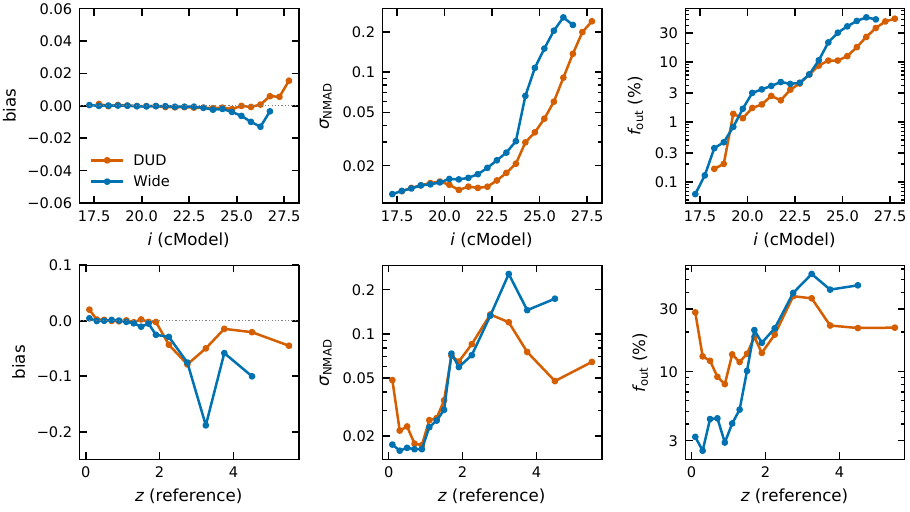}
\caption{Bias (left), $\snmad$ (middle) and outlier fraction (right) of all test galaxies and quasars against $i$ magnitude (top) and against the reference redshift (bottom), in DUD and Wide. Bins with at least 200 objects.} \label{fig:perf}
\end{figure*}

Figure~\ref{fig:perf} shows the bias, $\snmad$ and outlier fraction of all test galaxies and quasars against $i$ magnitude and against the reference redshift. In DUD, the bias stays within 0.002 up to $i = 26$ and reaches $+0.006$ to $+0.015$ beyond $i = 26.5$. The scatter is $\snmad = 0.013$--0.015 at $i < 23$. It rises to 0.018--0.021 at $23 < i < 24$, 0.030--0.035 at $24 < i < 25$, 0.045--0.060 at $25 < i < 26$, and 0.09--0.24 beyond. The outlier fraction is below 2\% at $i < 21$, 2--4\% at $21 < i < 23$, 6--10\% at $23 < i < 25$, 12--17\% at $25 < i < 26$, and 26--52\% beyond.

In Wide, the bias stays within 0.002 up to $i = 24.5$ and is $-0.003$ to $-0.013$ at fainter magnitudes. The scatter rises slowly from $\snmad = 0.012$ at $i \approx 17$ to 0.017 at $i \approx 22$ and 0.030 at $23.5 < i < 24$. Beyond $i = 24$, most reference redshifts are COSMOS-Web photo-$z$, and $\snmad$ jumps to 0.07--0.26. The outlier fraction is below 1\% at $i < 19.5$, 2--5\% at $20 < i < 23$, 6--11\% at $23 < i < 24$, and 21--55\% beyond.

As a function of the reference redshift, the bias in DUD stays within 0.003 at $0.2 < z < 2$. It is $+0.02$ at $z < 0.2$ and $-0.04$ to $-0.08$ at $2 < z < 3.5$. The scatter is $\snmad = 0.017$--0.027 at $0.2 < z < 1.4$, 0.065--0.085 at $1.6 < z < 2.5$, 0.12--0.14 at $2.5 < z < 3.5$, and 0.05--0.08 beyond. The outlier fraction is 8--14\% at $0.2 < z < 1.6$, 14--19\% at $1.6 < z < 2.5$, and 21--37\% beyond.

In Wide, the bias stays within 0.005 up to $z = 1.4$. It is $-0.005$ to $-0.03$ at $1.4 < z < 2.5$ and $-0.06$ to $-0.19$ beyond, where the Wide training set has few galaxies. The scatter is $\snmad = 0.016$--0.018 at $z < 1$, 0.023--0.030 at $1 < z < 1.6$, 0.06--0.07 at $1.6 < z < 2.5$, and 0.13--0.26 beyond. The outlier fraction is 2.5--5\% below $z = 1.4$, 10--21\% at $1.4 < z < 2.5$, and 40--55\% beyond.

\subsection{Calibration of the PDz} \label{sec:pit}
\begin{figure*}
\centering
\includegraphics[width=0.75\textwidth]{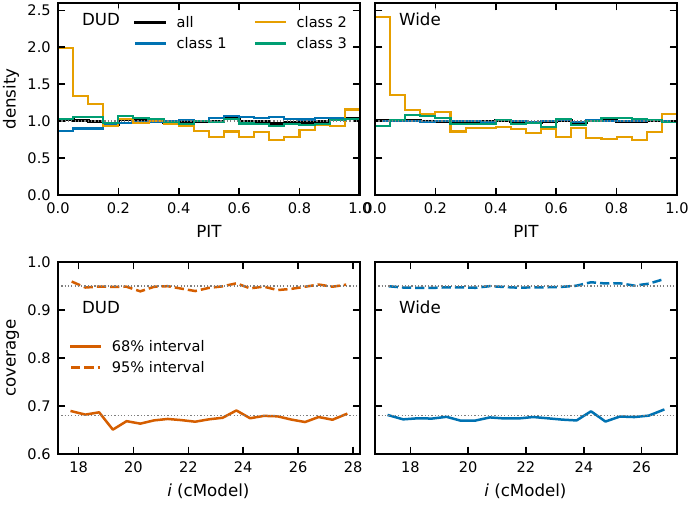}
\caption{Top: PIT distributions of the test set in DUD (left) and Wide (right), for all objects and for each class. A calibrated \pdz\ gives a flat distribution (dotted line). Bottom: fraction of reference redshifts inside the 68\% (solid) and 95\% (dashed) intervals against $i$ magnitude. Dotted lines mark the nominal values. Bins of 0.5\,mag with at least 200 objects. The statistical error of the coverage is up to 0.03 in the sparsest bins.} \label{fig:pit}
\end{figure*}

Table~\ref{tab:photoz} and Figure~\ref{fig:pit} show that the \pdz\ are well calibrated. Over the whole test set, the 68\% and 95\% intervals contain the reference redshift 67.3\% and 94.7\% of the time in DUD, and 67.4\% and 94.8\% of the time in Wide. The PIT distribution differs from uniform by at most 0.004. In every bin of $i$ magnitude in both layers, from $i < 20$ to $i > 26$, the coverage stays between 0.667 and 0.681 for the 68\% intervals and between 0.943 and 0.957 for the 95\% intervals.

The calibration is fitted on all classes together, so it holds for the mixture and not exactly for each class. The low-precision class-2 redshifts have larger errors than the class-1 redshifts that dominate the training set, and their 68\% coverage is only 0.60. The class-1 redshifts of DUD have a slightly high 68\% coverage of 0.685. The coverage is also lower at $z > 3$ (0.35--0.55). This is expected, because the PIT values of a sample selected by its true redshift need not be uniform.

\subsection{Redshift distributions} \label{sec:nz}
\begin{deluxetable*}{lrrrrrrrrr}
\tablecaption{Mean redshift of the stacked \pdz\ minus the mean reference redshift of the test set, in slices of $z_{\rm best}$. \label{tab:nz}}
\tablehead{\colhead{} & \colhead{0--0.3} & \colhead{0.3--0.6} & \colhead{0.6--0.9} & \colhead{0.9--1.2} & \colhead{1.2--1.5} & \colhead{1.5--2} & \colhead{2--3} & \colhead{3--4} & \colhead{4--6}}
\startdata
DUD  & $-0.009$ & $+0.002$ & $-0.005$ & $-0.002$ & $-0.003$ & $+0.015$ & $+0.003$ & $+0.027$ & $-0.058$ \\
Wide & $-0.001$ & $-0.003$ & $-0.002$ & $+0.000$ & $-0.000$ & $-0.009$ & $-0.007$ & $+0.006$ & $+0.121$ \\
\enddata
\tablecomments{All test galaxies and quasars of each layer, selected by their $z_{\rm best}$.}
\end{deluxetable*}

\begin{figure*}
\centering
\includegraphics[width=\textwidth]{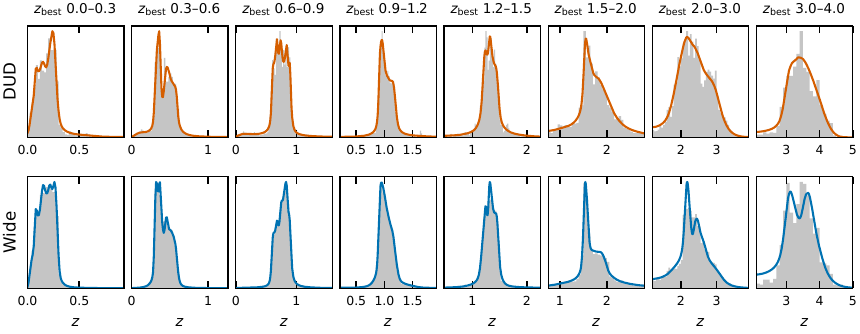}
\caption{Redshift distributions of the test galaxies and quasars in slices of $z_{\rm best}$ (columns), in DUD (top) and Wide (bottom). Gray: histogram of the reference redshifts. Line: stacked \pdz. Both are normalized to unit area.} \label{fig:nz}
\end{figure*}

For many applications, what matters is the redshift distribution of a sample selected by photo-$z$. I therefore select the test galaxies in slices of $z_{\rm best}$ and compare the mean of their stacked \pdz\ with the mean of their reference redshifts (Table~\ref{tab:nz}). In DUD, the two means agree within 0.009 below $z = 1.5$ and within 0.03 up to $z = 4$. In Wide, they agree within 0.004 in every slice below $z = 1.5$ and within 0.009 up to $z = 3$. The stacked distributions are slightly narrower than the true ones. Below $z = 3$, their standard deviation is smaller by up to 0.04 in DUD and by 0.004 to 0.018 in Wide. Figure~\ref{fig:nz} compares the stacked \pdz\ with the distributions of the reference redshifts.

\subsection{Ten-fold cross-validation in DUD} \label{sec:kfold}
\begin{deluxetable*}{llrcccccc}
\tablecaption{Ten-fold cross-validation of the photometric redshifts in DUD compared with the test set. \label{tab:kfold}}
\tablehead{\colhead{} & \colhead{} & \colhead{} & \multicolumn{2}{c}{$\snmad$} & \multicolumn{2}{c}{$\fout$} & \multicolumn{2}{c}{$C_{68}$} \\ \colhead{} & \colhead{Sample} & \colhead{$N_{\rm fold}$} & \colhead{Folds} & \colhead{Test} & \colhead{Folds} & \colhead{Test} & \colhead{Folds} & \colhead{Test}}
\startdata
              & all                & 58,134 & $0.0301 \pm 0.0004$ & 0.0300 & $0.152 \pm 0.001$ & 0.152 & $0.674 \pm 0.003$ & 0.673 \\
              & class 1            & 25,442 & $0.0146 \pm 0.0003$ & 0.0144 & $0.033 \pm 0.001$ & 0.034 & $0.684 \pm 0.003$ & 0.685 \\
              & class 2            &  4,197 & $0.0224 \pm 0.0005$ & 0.0226 & $0.067 \pm 0.003$ & 0.069 & $0.600 \pm 0.006$ & 0.595 \\
              & class 3            & 28,496 & $0.0803 \pm 0.0008$ & 0.0791 & $0.270 \pm 0.003$ & 0.268 & $0.676 \pm 0.003$ & 0.673 \\
\enddata
\tablecomments{Folds: mean and standard deviation of the ten folds, each predicted by a model trained on the other nine. $N_{\rm fold}$: mean number of objects per fold. Test: the DUD test set after training on all ten folds (Table~\ref{tab:photoz}).}
\end{deluxetable*}

Table~\ref{tab:kfold} compares the ten-fold cross-validation in DUD (Section~\ref{sec:design}) with the test set. The folds agree closely with one another. Over all galaxies and quasars, $\snmad$ ranges from 0.0294 to 0.0307 between the folds, with a standard deviation of 0.0004, and the outlier fraction ranges from 14.9\% to 15.4\%. The 68\% coverage is $0.674 \pm 0.003$. The 95\% coverage lies between 0.946 and 0.954 in every fold for all objects and for classes 1 and 3, and between 0.90 and 0.91 for class 2. The $\snmad$, outlier fraction and coverage of the test set agree with the folds within 1.5 standard deviations in every class, although the test-set model was trained on about 11\% more galaxies. The test-set results are therefore not a lucky draw, and they are not sensitive to modest changes in the training set.

\subsection{Stars} \label{sec:stars}
\begin{figure}
\centering
\includegraphics[width=\columnwidth]{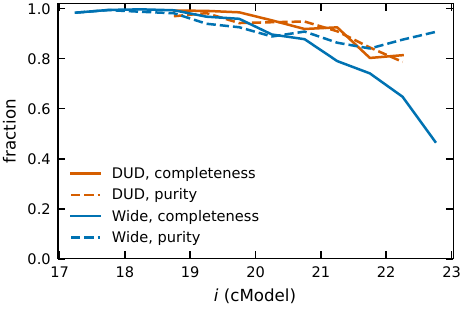}
\caption{Completeness (solid) and purity (dashed) of the stars selected with $P_{\rm STAR} > 0.5$ in the test set, against $i$ magnitude. Bins of 0.5\,mag with at least 50 objects.} \label{fig:stars}
\end{figure}

The star probability separates stars from galaxies and quasars with an area under the receiver operating characteristic curve of 0.999 in both layers. With $P_{\rm STAR} > 0.5$, 98.6\% of the stars brighter than $i = 20$ are selected, with a purity of 96--97\% in both layers. At $22 < i < 23$, where few spectroscopic stars exist, the completeness falls to 80\% in DUD and 59\% in Wide. Figure~\ref{fig:stars} shows the completeness and purity against magnitude.

\section{Predictions outside the training sample} \label{sec:od}

An empirical code is reliable where its training set is representative of the target galaxies. In practice, spectroscopic training sets are brighter than the target samples, and some types of galaxies are under-represented. \citet{Shen2024} show that engression can predict outside the range of its training data more reliably than a fit of a single value. To test how well this holds for photometric redshifts, I train the first-stage networks on one part of the DUD class-1 sample and predict the rest. In the first test (OD1), the training sample contains only bright galaxies, with $i < 22$, and the prediction sample only faint ones, with $i \geq 22$. Galaxies without a cModel $i$ measurement are left out of OD1, because their magnitude is unknown. In the second test (OD2), the training sample contains only blue galaxies, with a cModel color $g - r < 0.8$, and the prediction sample only red ones, with $g - r \geq 0.8$. In both tests, the networks never see a galaxy like those they predict. Both tests use only secure spectroscopic redshifts, so the truth is reliable. The training samples are drawn from the training set of Section~\ref{sec:design}, and the prediction samples from both the training and the test sets. The settings are those of the main DUD run, without the fast networks and the star classifier. I use the \pdz\ of the first-stage networks before calibration, because the calibration is fitted on the training galaxies and is not part of engression itself.

Both tests are extreme cases. Each removes a whole part of the parameter space from the training set with a sharp cut. In real training sets, bright and blue galaxies are over-represented, but fainter and redder galaxies are still present in smaller numbers, and there is no sharp edge. Real cases mix both effects, and each of them is much milder than in these tests.

\begin{figure*}
\centering
\includegraphics[width=\textwidth]{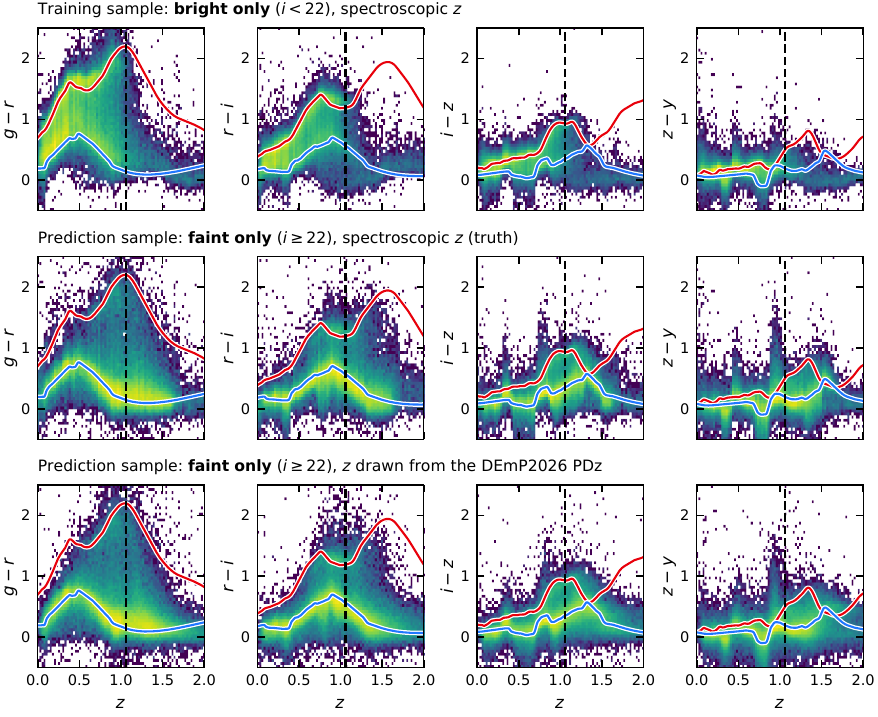}
\caption{Colors against redshift in the test OD1. Here the first-stage networks are trained only on bright galaxies, the DUD galaxies with secure spectroscopic redshifts and $i < 22$. Then they predict only faint galaxies, with $i \geq 22$. Columns: cModel $g - r$, $r - i$, $i - z$ and $z - y$. Top: the training sample at its spectroscopic redshift. Middle: the prediction sample at its spectroscopic redshift, which the networks never see. Bottom: the prediction sample with each galaxy placed at one redshift drawn at random from its \pdz, before calibration. A random draw is used instead of $z_{\rm best}$, because the drawn redshifts follow the full \pdz. If the \pdz\ are correct, the bottom row reproduces the middle row. Black dashed line: $z = 1.06$, below which 95\% of the training galaxies lie. The red and blue solid lines indicate the colors of the elliptical and irregular galaxy templates of \citet{Coleman1980}, as extended by \citet{Benitez2004}, computed with the HSC filter transmission curves \citep{Kawanomoto2018}. Color: number of galaxies per cell, on a logarithmic scale. The middle and bottom rows share one scale.} \label{fig:od1}
\end{figure*}

In OD1, the training sample contains 134,805 bright galaxies and the prediction sample 126,146 faint ones. Only 6.2\% of the training galaxies lie at $z > 1$, compared with 45.5\% of the prediction sample. Figure~\ref{fig:od1} shows the colors against redshift. When each galaxy of the prediction sample is placed at a redshift drawn from its \pdz\ (bottom row), the distribution reproduces that of the spectroscopic redshifts (middle row). This includes the features at $1 < z < 1.6$, where the training sample contains few galaxies, such as the track of the blue galaxies in $r - i$ and $i - z$ and the narrow features in $z - y$. For the OD1 prediction sample, $z_{\rm best}$ gives $\snmad = 0.021$ and $\fout = 9.6\%$. For comparison, the galaxies of the test set with secure spectroscopic redshifts and $i \geq 22$, for which the training set covers all magnitudes, give $\snmad = 0.015$ and $\fout = 5.0\%$. At $1 < z < 1.5$, where the training sample has 5,325 galaxies and the prediction sample 45,267, $\snmad$ is 0.023 and the bias is $+0.005$. The prediction degrades at $1.5 < z < 2$, with $\snmad = 0.068$ and $\fout = 21\%$. It fails beyond $z = 2$, where the training sample contains only 1,571 galaxies. There, the bias is $-0.61$, and 81\% of the galaxies are outliers. These galaxies make up 4\% of the prediction sample.

\begin{deluxetable*}{llrrrrrrrrrr}
\tablecaption{Photometric redshifts of the prediction samples in the tests OD1 and OD2, from the \pdz\ of the first-stage networks before calibration. \label{tab:od}}
\tablehead{\colhead{Test} & \colhead{$z$} & \colhead{$N_{\rm train}$} & \colhead{$N_{\rm pred}$} & \colhead{Bias} & \colhead{$\snmad$} & \colhead{$\fout$} & \colhead{$C_{68}$} & \colhead{$C_{95}$} & \colhead{CRPS} & \colhead{$\langle z \rangle_{\rm PDz}$} & \colhead{$\langle z \rangle_{\rm spec}$}}
\startdata
OD1 & all      & 134,805 & 126,146 & $+0.0018$ & 0.0206 & 0.096 & 0.642 & 0.864 & 0.138 & 0.943 & 1.005 \\
    & 0--1     & 126,382 &  68,700 & $+0.0015$ & 0.0158 & 0.057 & 0.678 & 0.896 & 0.049 & \nodata & \nodata \\
    & 1--1.5   &   5,325 &  45,267 & $+0.0051$ & 0.0226 & 0.052 & 0.664 & 0.906 & 0.064 & \nodata & \nodata \\
    & 1.5--2   &   1,527 &   6,864 & $-0.0003$ & 0.0683 & 0.212 & 0.542 & 0.778 & 0.216 & \nodata & \nodata \\
    & $>2$     &   1,571 &   5,315 & $-0.6078$ & 0.1201 & 0.809 & 0.124 & 0.199 & 1.826 & \nodata & \nodata \\
OD2 & all      & 156,125 & 101,743 & $-0.0119$ & 0.0331 & 0.039 & 0.633 & 0.906 & 0.068 & 0.542 & 0.591 \\
    & 0--0.5   &  47,733 &  50,470 & $-0.0047$ & 0.0302 & 0.018 & 0.696 & 0.938 & 0.036 & \nodata & \nodata \\
    & 0.5--1   &  54,970 &  44,050 & $-0.0212$ & 0.0359 & 0.034 & 0.573 & 0.891 & 0.051 & \nodata & \nodata \\
    & 1--1.5   &  40,964 &   5,523 & $-0.0089$ & 0.0263 & 0.065 & 0.678 & 0.907 & 0.072 & \nodata & \nodata \\
    & $>1.5$   &  12,458 &   1,700 & $-0.3398$ & 0.3576 & 0.708 & 0.171 & 0.319 & 1.430 & \nodata & \nodata \\
\enddata
\tablecomments{The prediction samples are the DUD class-1 galaxies with $i \geq 22$ (OD1) and with $g - r \geq 0.8$ (OD2). $z$: slice of spectroscopic redshift. $N_{\rm train}$, $N_{\rm pred}$: number of training and prediction galaxies in the slice. Bias, $\snmad$ and $\fout$ use the $z_{\rm best}$ of each \pdz. $\langle z \rangle_{\rm PDz}$, $\langle z \rangle_{\rm spec}$: mean redshift of the stacked \pdz\ and of the spectroscopic redshifts. They are given for the whole prediction samples only, because the \pdz\ of galaxies selected by their true redshift extend beyond the slice.}
\end{deluxetable*}

\begin{figure*}
\centering
\includegraphics[width=\textwidth]{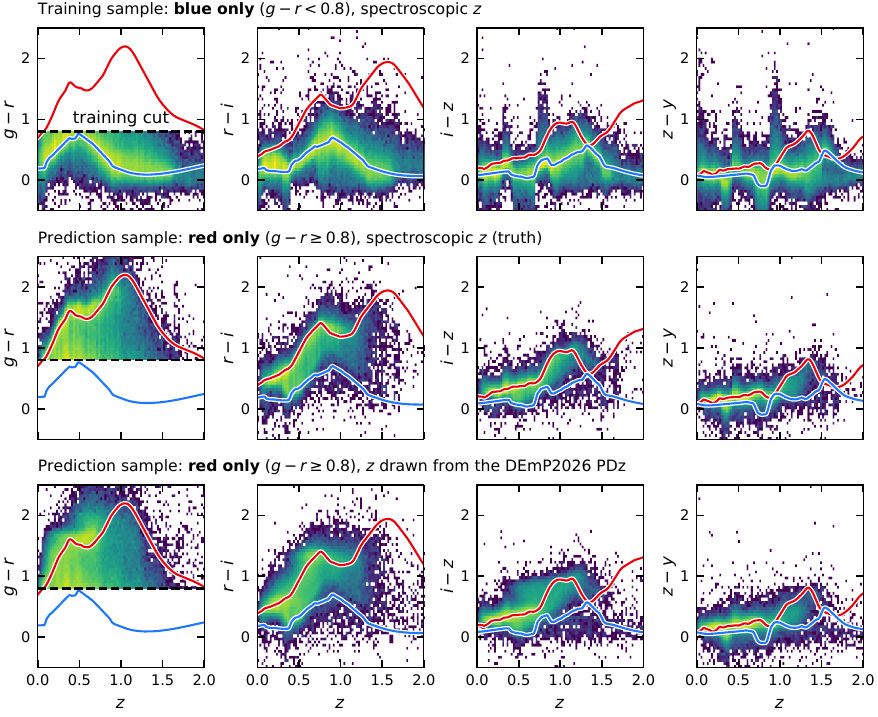}
\caption{As Figure~\ref{fig:od1}, for the test OD2. Here the networks are trained only on blue galaxies, with cModel $g - r < 0.8$. Then they predict only red galaxies, with $g - r \geq 0.8$. Black dashed line in the $g - r$ panels: the cut at $g - r = 0.8$ that separates the two samples. Because of this cut, the training sample lies below the line and the prediction sample above it.} \label{fig:od2}
\end{figure*}

In OD2, the training sample contains only blue galaxies, 156,125 of them, and the prediction sample only red ones, 101,743 of them. The redshift ranges of the two samples overlap, but the colors of the prediction sample lie outside those of the training sample. Figure~\ref{fig:od2} shows that the predicted distribution reproduces the overall relations between color and redshift. However, the red sequence at $0.6 < z < 1$ is placed at slightly too low redshift. For the prediction sample, $\snmad = 0.033$, $\fout = 3.9\%$ and the bias is $-0.012$. At $0.5 < z < 1$, the bias is $-0.021$. The red galaxies at $z > 1.5$ make up 1.7\% of the prediction sample, and most of them lie at $z > 3$. For these galaxies, the prediction fails. Table~\ref{tab:od} gives the statistics of both tests.

\begin{figure}
\centering
\includegraphics[width=\columnwidth]{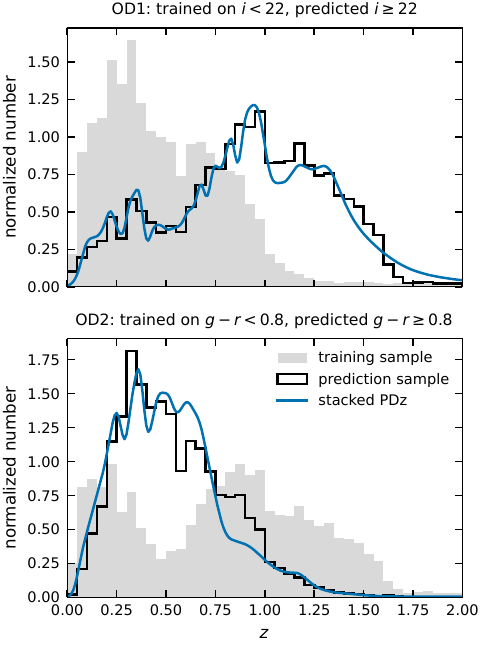}
\caption{Redshift distributions in the tests OD1 (top) and OD2 (bottom). Gray: training sample. Black: spectroscopic redshifts of the prediction sample. Blue: stacked \pdz\ of the prediction sample, before calibration. All are normalized to unit area.} \label{fig:odnz}
\end{figure}

Figure~\ref{fig:odnz} compares the stacked \pdz\ of the two prediction samples with the distributions of their spectroscopic redshifts. In OD1, the stacked \pdz\ follows the redshift distribution of the prediction sample up to $z \approx 1.6$, although most training galaxies lie at $z < 1$. The galaxies at $z > 2$ are placed at lower redshifts, so the mean redshift of the stacked \pdz\ is 0.943, compared with 1.005 (Table~\ref{tab:od}). In OD2, the stacked \pdz\ reproduces the peak of the redshift distribution at $0.25 < z < 0.5$. However, it places too many galaxies at $0.55 < z < 0.7$ and too few at $0.75 < z < 1$, because the red sequence is placed at slightly too low redshift. The mean redshift of the stacked \pdz\ is 0.542, compared with 0.591.

In both tests, the \pdz\ before calibration are slightly too narrow. Their 68\% intervals contain the spectroscopic redshift for 64.2\% (OD1) and 63.3\% (OD2) of the prediction sample (Table~\ref{tab:od}). The calibration of Section~\ref{sec:calibration}, fitted on the training sample, changes this only a little, to 0.63 and 0.65. The \pdz\ of the catalog are too broad, however, with 68\% coverages of 0.73 and 0.75. The broadening comes from the correction for galaxies far from the training set (Section~\ref{sec:ood}). The galaxies of the prediction samples lie much farther from the training set than the training galaxies themselves, with a median distance of $r = 3.2$ (OD1) and 3.6 (OD2), compared with 1 for the training galaxies. The correction at such distances is fitted on the few training galaxies that lie there. These are mostly unusual objects with poor estimates, whereas the prediction galaxies are ordinary galaxies that are distant only because of the cut. The correction is therefore too large for them. In summary, trained only on bright or only on blue galaxies, the networks still predict faint or red galaxies well. They extrapolate usefully into regions of color, magnitude and redshift that the training set covers only sparsely.

\section{Comparison with the original DEmP} \label{sec:orig}
The original DEmP photo-$z$ of PDR3 are public for every object. I compare the two codes on all test galaxies and quasars that have an original photo-$z$, using the same objects for both codes. The original training set is not public, so some test objects may have been part of it. For these objects, the original estimates are in-sample, which can only favor the original code. The improvements found here are therefore conservative. In DUD, the original run used an earlier processing of the same data with different object ids, and the original photo-$z$ are matched by position within 0.3\arcsec. In Wide, they are matched by object id.

\begin{deluxetable*}{llrrrrrrrrr}
\tablecaption{\demp\ and the original DEmP photo-$z$ of PDR3 on the same test objects. \label{tab:orig}}
\tablehead{\colhead{} & \colhead{} & \colhead{} & \multicolumn{4}{c}{\demp} & \multicolumn{4}{c}{Original DEmP} \\ \colhead{Layer} & \colhead{Sample} & \colhead{$N$} & \colhead{$\snmad$} & \colhead{$\fout$} & \colhead{$C_{68}$} & \colhead{CRPS} & \colhead{$\snmad$} & \colhead{$\fout$} & \colhead{$C_{68}$} & \colhead{CRPS}}
\startdata
DUD  & all     &  64,587 & 0.0300 & 0.152 & 0.673 & 0.179 & 0.0458 & 0.235 & 0.622 & 0.306 \\
     & class 1 &  28,194 & 0.0144 & 0.034 & 0.685 & 0.044 & 0.0206 & 0.055 & 0.707 & 0.070 \\
     & class 2 &   4,632 & 0.0226 & 0.069 & 0.595 & 0.054 & 0.0265 & 0.083 & 0.570 & 0.070 \\
     & class 3 &  31,761 & 0.0790 & 0.268 & 0.673 & 0.317 & 0.1571 & 0.416 & 0.555 & 0.549 \\
Wide & all     & 402,193 & 0.0191 & 0.052 & 0.674 & 0.061 & 0.0277 & 0.108 & 0.551 & 0.111 \\
     & class 1 & 379,897 & 0.0181 & 0.038 & 0.675 & 0.046 & 0.0263 & 0.093 & 0.553 & 0.092 \\
     & class 2 &   5,519 & 0.0258 & 0.068 & 0.601 & 0.060 & 0.0286 & 0.083 & 0.500 & 0.074 \\
     & class 3 &  16,777 & 0.1331 & 0.365 & 0.676 & 0.396 & 0.1949 & 0.461 & 0.507 & 0.559 \\
\enddata
\tablecomments{All test galaxies and quasars with an original \pdz. The 40 DUD objects without an original \pdz\ (9 without a match within 0.3\arcsec\ and 31 whose original \pdz\ is zero at all redshifts) and the 2,719 Wide objects whose original \pdz\ is zero at all redshifts are left out. The original \pdz\ is given on a grid of 0.01 in $z$.}
\end{deluxetable*}

\begin{deluxetable*}{lrrrrrrrrr}
\tablecaption{Mean redshift of the stacked \pdz\ minus the mean reference redshift on the comparison sample, in slices of $z_{\rm best}$. \label{tab:orignz}}
\tablehead{\colhead{} & \colhead{0--0.3} & \colhead{0.3--0.6} & \colhead{0.6--0.9} & \colhead{0.9--1.2} & \colhead{1.2--1.5} & \colhead{1.5--2} & \colhead{2--3} & \colhead{3--4} & \colhead{4--6}}
\startdata
DUD, \demp\           & $-0.009$ & $+0.002$ & $-0.004$ & $-0.002$ & $-0.003$ & $+0.015$ & $+0.003$ & $+0.027$ & $-0.058$ \\
DUD, original DEmP  & $+0.003$ & $+0.010$ & $+0.035$ & $+0.053$ & $+0.071$ & $+0.149$ & $+0.631$ & $+1.187$ & $+0.820$ \\
Wide, \demp\          & $-0.001$ & $-0.003$ & $-0.002$ & $+0.000$ & $-0.000$ & $-0.008$ & $-0.007$ & $+0.006$ & $+0.121$ \\
Wide, original DEmP & $-0.018$ & $-0.043$ & $-0.051$ & $-0.024$ & $-0.009$ & $+0.110$ & $+0.386$ & $+1.234$ & $+1.788$ \\
\enddata
\tablecomments{The objects of Table~\ref{tab:orig}. Each code is sliced by its own $z_{\rm best}$.}
\end{deluxetable*}

\begin{figure*}
\centering
\includegraphics[width=0.72\textwidth]{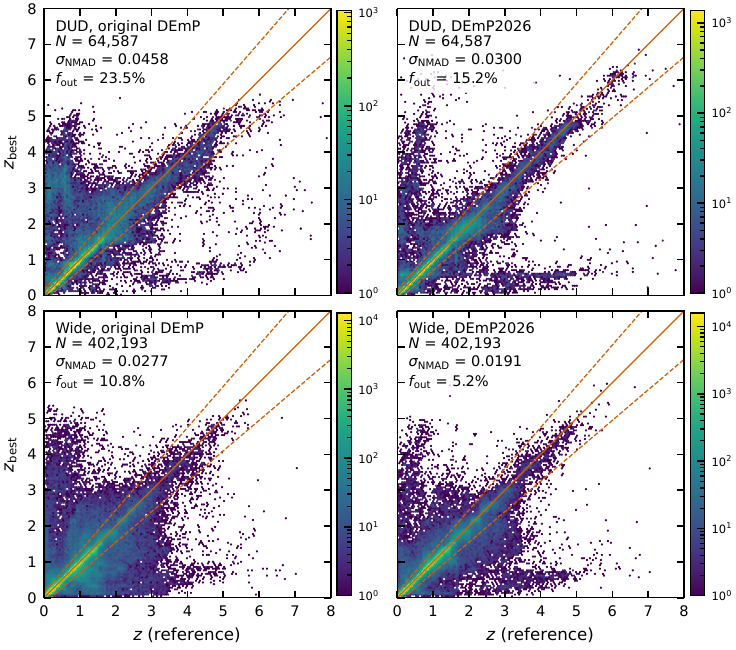}
\caption{$z_{\rm best}$ against the reference redshift for the original DEmP (left) and \demp\ (right), on the same test objects (all test galaxies and quasars with an original photo-$z$), in DUD (top) and Wide (bottom). This leaves out 40 DUD and 2,719 Wide objects of Figure~\ref{fig:zz} (Table~\ref{tab:orig}). Colors and lines as in Figure~\ref{fig:zz}.} \label{fig:origzz}
\end{figure*}

\begin{figure*}
\centering
\includegraphics[width=0.75\textwidth]{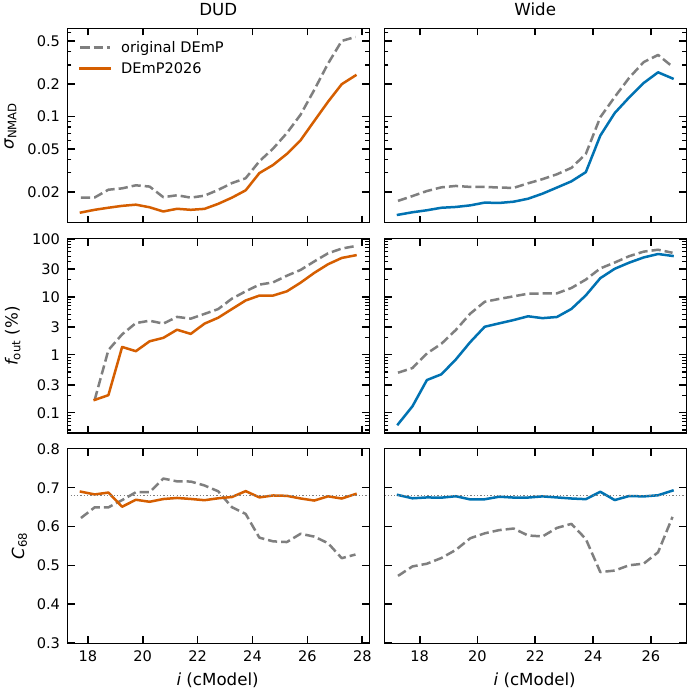}
\caption{$\snmad$, outlier fraction and 68\% coverage against $i$ magnitude for the original DEmP (dashed) and \demp\ (solid) on the comparison sample, in DUD (left) and Wide (right). The dotted line marks the nominal coverage. Bins of 0.5\,mag with at least 200 objects.} \label{fig:origmag}
\end{figure*}

Table~\ref{tab:orig} shows the comparison, and Figure~\ref{fig:origzz} shows $z_{\rm best}$ against the reference redshift for both codes. \demp\ has a lower $\snmad$, outlier fraction and CRPS in every class of both layers. Over all objects, it reduces $\snmad$ by 35\% in DUD and 31\% in Wide, the outlier fraction by 35\% and 52\%, and the CRPS by 41\% and 45\%. The largest gain is in the calibration of the \pdz. The 68\% intervals of the original DEmP contain the reference redshift only 62\% (DUD) and 55\% (Wide) of the time, while those of \demp\ contain it 67\% of the time. The gain holds in every bin of $i$ magnitude, with reductions in $\snmad$ of 24--57\% in DUD and 26--35\% in Wide (Figure~\ref{fig:origmag}).

The gain is also clear in the redshift distributions. I measure them as in Section~\ref{sec:nz}, with each code sliced by its own $z_{\rm best}$ (Table~\ref{tab:orignz}). Below $z = 1.5$, the stacked \pdz\ of the original code are offset by up to 0.07 in DUD and by up to 0.05 in Wide, compared with at most 0.009 and 0.003 for \demp. The original code also places too many galaxies at high $z_{\rm best}$.

These differences do not come from the codes alone, because the two runs were trained on different reference redshifts. The \demp\ training set is dominated by COSMOS-Web in DUD and by DESI DR1 in Wide. Both were released in 2025, after the original photo-$z$ of PDR3 were produced, so neither can have been part of the original training set. The two runs also differ in their input quantities and, in DUD, in the processing of the photometry (Section~\ref{sec:hsc}). Part of the gain therefore reflects the newer reference redshifts, in particular for the COSMOS-Web galaxies.

To gauge how large this part is, I split the class-1 test objects by whether their redshift includes a DESI DR1 measurement. In DUD, \demp\ reduces $\snmad$ by 33\% and the outlier fraction by 42\% for the 21,542 objects with DESI, and by 21\% and 26\% for the 6,652 objects without DESI. In Wide, the reductions are 33\% and 63\% for the 326,095 objects with DESI, and 20\% and 36\% for the 53,802 objects without DESI. The gain is smaller for the objects without DESI, but it remains substantial, even though some of them may have been in the original training set. Separating the effects of the code and of the training data exactly would require training both codes on the same data, which is beyond the scope of this paper.

\section{Stellar masses and star-formation rates} \label{sec:mass}
\begin{deluxetable*}{llrrrrrrrr}
\tablecaption{Stellar masses and star-formation rates of the test set. \label{tab:mass}}
\tablehead{\colhead{Quantity} & \colhead{Reference, layer} & \colhead{$N$} & \colhead{Bias} & \colhead{$\snmad$} & \colhead{$\sigma$} & \colhead{$\fout$} & \colhead{$f_{>0.5}$} & \colhead{$C_{68}$} & \colhead{$C_{95}$}}
\startdata
$\log M_*$   & DESI, DUD         &  12,411 & $-0.001$ & 0.127 & 0.255 & 0.123 & 0.049 & 0.679 & 0.948 \\
             & DESI, Wide        & 264,458 & $+0.001$ & 0.144 & 0.286 & 0.153 & 0.062 & 0.681 & 0.950 \\
             & COSMOS-Web, DUD   &  30,049 & $+0.016$ & 0.255 & 0.590 & 0.320 & 0.186 & 0.679 & 0.951 \\
             & COSMOS-Web, Wide  &  16,664 & $+0.004$ & 0.333 & 0.662 & 0.404 & 0.255 & 0.676 & 0.948 \\
$\log$\,SFR  & DESI, DUD         &  12,411 & $-0.005$ & 0.181 & 0.564 & 0.264 & 0.164 & 0.679 & 0.949 \\
             & DESI, Wide        & 264,458 & $-0.005$ & 0.188 & 0.582 & 0.277 & 0.170 & 0.681 & 0.949 \\
             & COSMOS-Web, DUD   &  30,049 & $+0.029$ & 0.275 & 0.842 & 0.336 & 0.215 & 0.678 & 0.949 \\
             & COSMOS-Web, Wide  &  16,664 & $+0.027$ & 0.412 & 1.010 & 0.469 & 0.305 & 0.672 & 0.948 \\
\enddata
\tablecomments{Differences are the best estimate minus the reference value, in dex. $\sigma$: standard deviation. $\fout$: outlier fraction, the fraction with an absolute difference larger than 0.3\,dex. $f_{>0.5}$: fraction with an absolute difference larger than 0.5\,dex.}
\end{deluxetable*}

\begin{figure*}
\centering
\includegraphics[width=\textwidth]{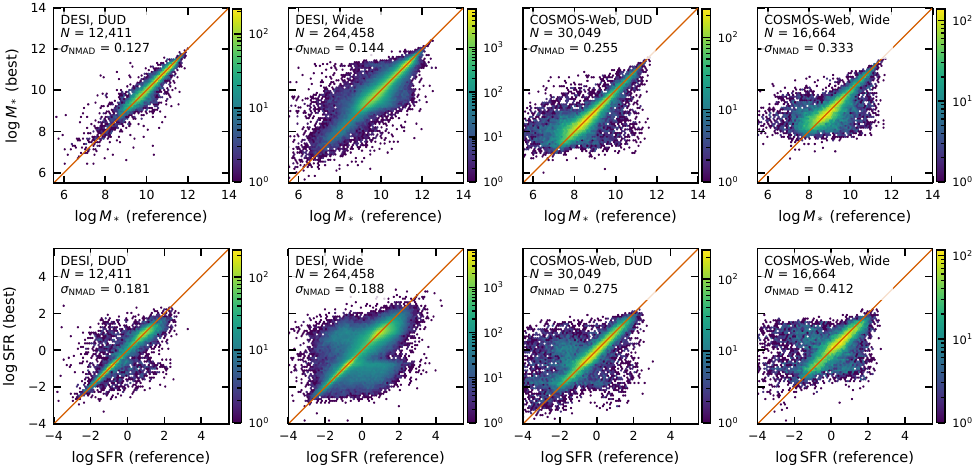}
\caption{Best estimate against reference value for $\log M_*$ (top) and $\log$\,SFR (bottom) in the test sets, for the DESI and COSMOS-Web reference values in DUD and Wide. Color: number of objects per hexagonal cell, on a logarithmic scale (empty cells are white). The line marks equality. For display, the best estimates, which lie on a grid of step 0.01\,dex, are spread uniformly within their grid step.} \label{fig:mass}
\end{figure*}

Table~\ref{tab:mass} gives the stellar masses and SFRs of the test set, and Figure~\ref{fig:mass} compares the best estimates with the reference values. For each galaxy, the difference is the best estimate minus the reference value, in dex. As for the redshifts, the bias is the median difference, and $\snmad$ is computed from the differences in the same way. The outlier fraction $\fout$ is the fraction of galaxies with a difference larger than 0.3\,dex in absolute value. For the DESI reference values, $\log M_*$ is recovered with $\snmad = 0.127$\,dex in DUD and 0.144\,dex in Wide, and the outlier fractions are 12\% and 15\%. For galaxies brighter than $i = 22$, the scatter is 0.09--0.12\,dex. It is smallest, 0.08--0.09\,dex, for the massive galaxies ($10.5 < \log M_* < 11.5$) that dominate the DESI sample.

The COSMOS-Web reference values reach much fainter galaxies, to $i \approx 27$. For them, $\log M_*$ is recovered with $\snmad = 0.255$\,dex in DUD and 0.333\,dex in Wide, and the outlier fractions are 32\% and 40\%. At $z < 0.3$, the stellar masses are overestimated by 1.0\,dex in DUD and 1.3\,dex in Wide (1,907 and 871 galaxies). This reflects a degeneracy of the HSC photometry, not a failure of the mass estimate. Almost all of these galaxies are faint (median $i = 26.3$ in DUD), and their reference redshifts are COSMOS-Web photo-$z$. For 61\% (DUD) and 66\% (Wide) of them, the \demp\ photo-$z$ disagrees with the reference redshift, with a median $z_{\rm best}$ of 0.9--1.0. For these galaxies, the stellar mass is high by 1.6--1.7\,dex, as expected for the same photometry placed at the higher redshift. Where the photo-$z$ agrees with the reference redshift, the bias is 0.2--0.3\,dex. For the 257 galaxies with a spectroscopic redshift below 0.3, it is 0.04\,dex (DUD) and 0.16\,dex (Wide). Such low-redshift galaxies make up only 4--6\% of the faint ($i > 25$) COSMOS-Web sample, so the mass estimate favors the far more common higher-redshift solution. If the galaxies are instead binned by the \demp\ photo-$z$, the mass bias is below 0.06\,dex in DUD and 0.11\,dex in Wide in every slice. The \pdz\ of these galaxies keeps the low-redshift solution, and its 95\% interval reaches $z < 0.3$ for 91--94\% of them. The probability distribution of the stellar mass is narrower, and its 95\% interval contains the reference value for only 64--69\% of them.

For the DESI reference values, $\log{\rm SFR}$ is recovered with $\snmad = 0.181$\,dex in DUD and 0.188\,dex in Wide, and the outlier fractions are 26\% and 28\%. The SFRs have broader, non-Gaussian errors, with standard deviations of 0.56 and 0.58\,dex. In both layers, a third of the DESI reference SFRs are at the floor of Section~\ref{sec:masslabels}. For these galaxies, \demp\ gives $\snmad = 0.08$\,dex in both layers. For the others, it gives 0.29\,dex in DUD and 0.30\,dex in Wide. For the most strongly star-forming galaxies ($\log{\rm SFR} > 2$), the estimates are low by 0.74\,dex in DUD and 0.72\,dex in Wide. This is a regression toward the mean of the training set, and the intervals do not fully capture it, with $C_{68} = 0.18$ and 0.22.

For the COSMOS-Web reference values, $\log{\rm SFR}$ is recovered with $\snmad = 0.275$\,dex in DUD and 0.412\,dex in Wide, and the outlier fractions are 34\% and 47\%.

\begin{figure*}
\centering
\includegraphics[width=\textwidth]{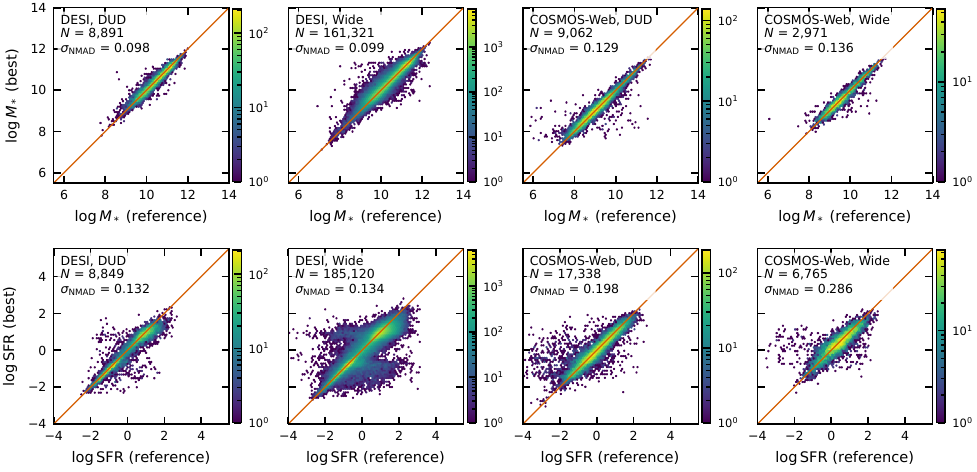}
\caption{As Figure~\ref{fig:mass}, for the galaxies whose 68\% interval has a half-width below 0.2\,dex in $\log M_*$ (top) or below 0.4\,dex in $\log$\,SFR (bottom).} \label{fig:masscut}
\end{figure*}

In every case, for both quantities and both reference catalogs, the 68\% and 95\% intervals contain the reference value 67--68\% and 95\% of the time. A large part of the scatter comes from galaxies with broad probability distributions, which the catalog identifies through the widths of their intervals. Figure~\ref{fig:masscut} shows the comparison of Figure~\ref{fig:mass} for the galaxies whose 68\% interval has a half-width below 0.2\,dex in $\log M_*$ or below 0.4\,dex in $\log{\rm SFR}$. For the DESI stellar masses, this cut keeps 72\% (DUD) and 61\% (Wide) of the galaxies. It reduces $\snmad$ from 0.127 to 0.098\,dex and from 0.144 to 0.099\,dex, and the outlier fraction from 12\% to 5\% and from 15\% to 5\%. For the COSMOS-Web stellar masses, it keeps 30\% (DUD) and 18\% (Wide) of the galaxies, with $\snmad = 0.13$ and 0.14\,dex. For the SFRs, it keeps 71\% (DUD) and 70\% (Wide) of the DESI galaxies with $\snmad = 0.13$\,dex, and 58\% and 41\% of the COSMOS-Web galaxies with $\snmad = 0.20$ and 0.29\,dex. A cut in magnitude is much less effective. For example, $i < 22$ keeps 75\% (DUD) and 73\% (Wide) of the DESI galaxies, but it reduces $\snmad$ of $\log M_*$ only to 0.107 and 0.119\,dex, and it removes 95--97\% of the COSMOS-Web galaxies.

As for the photometric redshifts (Section~\ref{sec:kfold}), I run a ten-fold cross-validation of the stellar masses and SFRs in DUD (Table~\ref{tab:kfoldmass}). The standard deviation of $\snmad$ between the folds is 0.001--0.003\,dex. The $\snmad$ of the test set is within 0.003\,dex of the mean of the folds, and the intervals are equally well calibrated in every fold.

\begin{deluxetable*}{llrcccccc}
\tablecaption{Ten-fold cross-validation of the stellar masses and star-formation rates in DUD compared with the test set. \label{tab:kfoldmass}}
\tablehead{\colhead{} & \colhead{} & \colhead{} & \multicolumn{2}{c}{$\snmad$ (dex)} & \multicolumn{2}{c}{$\fout$} & \multicolumn{2}{c}{$C_{68}$} \\ \colhead{Quantity} & \colhead{Reference} & \colhead{$N_{\rm fold}$} & \colhead{Folds} & \colhead{Test} & \colhead{Folds} & \colhead{Test} & \colhead{Folds} & \colhead{Test}}
\startdata
$\log M_*$    & DESI               & 11,095 & $0.129 \pm 0.001$   & 0.127  & $0.124 \pm 0.003$ & 0.123 & $0.679 \pm 0.005$ & 0.679 \\
              & COSMOS-Web         & 26,954 & $0.257 \pm 0.001$   & 0.255  & $0.319 \pm 0.003$ & 0.320 & $0.680 \pm 0.002$ & 0.679 \\
$\log$\,SFR   & DESI               & 11,095 & $0.182 \pm 0.002$   & 0.181  & $0.267 \pm 0.003$ & 0.264 & $0.679 \pm 0.005$ & 0.679 \\
              & COSMOS-Web         & 26,954 & $0.273 \pm 0.003$   & 0.275  & $0.334 \pm 0.003$ & 0.336 & $0.680 \pm 0.002$ & 0.678 \\
\enddata
\tablecomments{As Table~\ref{tab:kfold}, with the test values from Table~\ref{tab:mass}. $\fout$: outlier fraction (Table~\ref{tab:mass}).}
\end{deluxetable*}

\section{Computational cost} \label{sec:compute}
\begin{deluxetable}{lrrrrr}
\tablecaption{Run time and storage on one node with ten AMD Instinct\texttrademark\ MI210 GPUs. \label{tab:cost}}
\tablehead{\colhead{Run} & \colhead{$N_{\rm train}$} & \colhead{$N_{\rm target}$} & \colhead{Networks} & \colhead{Total} & \colhead{Models} \\ \colhead{} & \colhead{(M)} & \colhead{(M)} & \colhead{(h)} & \colhead{(h)} & \colhead{(GB)}}
\startdata
photo-$z$ DUD              & 0.58 & 0.07 & 0.59 & 2.09 & 1.76 \\
photo-$z$ Wide             & 3.64 & 0.49 & 3.62 & 7.88 & 2.10 \\
$M_*$ DESI DUD             & 0.11 & 0.06 & 0.12 & 0.20 & 0.90 \\
$M_*$ COSMOS-Web DUD       & 0.27 & 0.06 & 0.28 & 0.43 & 0.92 \\
$M_*$ DESI Wide            & 2.38 & 0.40 & 2.40 & 3.47 & 1.18 \\
$M_*$ COSMOS-Web Wide      & 0.15 & 0.40 & 0.16 & 0.42 & 0.90 \\
SFR DESI DUD               & 0.11 & 0.06 & 0.12 & 0.22 & 0.90 \\
SFR COSMOS-Web DUD         & 0.27 & 0.06 & 0.28 & 0.44 & 0.92 \\
SFR DESI Wide              & 2.38 & 0.40 & 2.39 & 3.99 & 1.18 \\
SFR COSMOS-Web Wide        & 0.15 & 0.40 & 0.16 & 0.44 & 0.90 \\
\enddata
\tablecomments{Networks: training of the 30 first-stage networks. Total: wall time of the whole run, including the calibration, the fast networks and the catalog of the target objects. The fast networks are used for the photo-$z$ only and take 0.9\,h in DUD and 1.7\,h in Wide, with 3.1 and 3.7 million galaxies without redshifts.}
\end{deluxetable}

Table~\ref{tab:cost} lists the cost of the runs. Each run used one node with ten AMD Instinct\texttrademark\ MI210 GPUs. The training time of the first-stage networks grows linearly with the size of the training set. It is 0.59\,h for the 0.58 million DUD galaxies and 3.6\,h for the 3.64 million Wide galaxies, a ratio of 6.1 for a ratio of 6.3 in size. The mass runs follow the same relation. The complete Wide photo-$z$ run takes 7.9\,h. This includes the calibration, the fast networks, which are also trained on 3.7 million galaxies without redshifts, and the catalog of 0.49 million test objects. The trained models occupy 2.1\,GB.

Each \pdz\ takes 60 bytes, or 68 bytes with the object id. The \pdz\ of the 0.49 million Wide test objects therefore occupy 34\,MB. The roughly 360 million primary Wide objects with $i < 26$, a number estimated from the random sample without redshifts, would need about 25\,GB. Once the models are trained, the catalog step computes the \pdz, point estimates, distances to the training set and $P_{\rm STAR}$. It took 11 minutes for the 0.49 million Wide test objects on one GPU, or 22 minutes per million objects. On a CPU node with AMD EPYC\texttrademark\ 9755 processors, a single process, using on average 2.3 cores, needs 53\,s for 20,000 objects, or 44 minutes per million. The objects are independent, so this step can be split over as many GPUs or CPU processes as are available.

\section{Summary} \label{sec:summary}
I have presented \demp, an empirical code for the probability distributions of the photometric redshift, stellar mass and SFR of galaxies. A set of neural networks that turn random numbers into possible redshifts, trained with the energy score (engression), learns the \pdz\ without assuming its shape. A calibration step, based on out-of-sample predictions for the training galaxies, corrects the \pdz\ so that their intervals contain the true redshift as often as stated. Faster networks, trained also on millions of galaxies without redshifts, then reproduce the calibrated \pdz\ in a single evaluation and store each \pdz\ as 30 quantiles in 60 bytes. I applied \demp\ to HSC PDR3 with public reference data only. On a random test set that was not used for any choice of the method, \demp\ gives the following results.
\begin{enumerate}
\item For the secure spectroscopic redshifts, $\snmad = 0.014$ and $\fout = 3.4\%$ in DUD, and $\snmad = 0.018$ and $\fout = 3.9\%$ in Wide. For the COSMOS-Web galaxies of DUD, which reach $i \approx 27$, $\snmad = 0.079$.
\item The \pdz\ are calibrated. The 68\% and 95\% intervals contain the reference redshift 67.3\% and 94.7\% of the time in DUD, and 67.4\% and 94.8\% of the time in Wide. In every magnitude bin, the coverages stay within 0.667--0.681 and 0.943--0.957.
\item Below $z = 1.5$, the mean redshift of galaxies selected in slices of $z_{\rm best}$ is recovered from the stacked \pdz\ to within 0.009 in DUD and 0.004 in Wide.
\item A ten-fold cross-validation in DUD gives a scatter of 0.0004 in $\snmad$ between the folds and agrees with the test set.
\item Trained only on bright galaxies ($i < 22$), the first-stage networks predict only faint galaxies ($i \geq 22$) with $\snmad = 0.021$ and recover their redshift distribution up to $z \approx 1.6$. Trained only on blue galaxies ($g - r < 0.8$), they predict only red galaxies ($g - r \geq 0.8$) with $\snmad = 0.033$.
\item On the same test objects, \demp\ reduces the $\snmad$ of the original DEmP photo-$z$ of PDR3 by 35\% (DUD) and 31\% (Wide), the outlier fraction by 35\% and 52\%, and the CRPS by 41\% and 45\%. It raises the 68\% coverage from 62\% and 55\% to 67\%. Part of this gain reflects the newer reference redshifts. However, for objects whose redshifts do not come from DESI DR1, $\snmad$ is still reduced by 20--21\%.
\item The same code gives $\log M_*$ with $\snmad = 0.13$--0.14\,dex for the DESI reference values and 0.26--0.33\,dex for the fainter COSMOS-Web reference values. It gives $\log$\,SFR with 0.18--0.19 and 0.28--0.41\,dex. The intervals are calibrated in every case.
\item The training time grows linearly with the size of the training set. The full Wide photo-$z$ run on 3.6 million training galaxies takes 7.9\,h on one node with ten GPUs, and the catalog step takes 22 minutes per million objects on one GPU.
\end{enumerate}
The calibration holds for the mixture of classes on which it is fitted, but not exactly for every class. Low-precision redshifts and quasars have lower coverage, and at $z > 3$ the Wide training set is too sparse for good point estimates. As for any empirical method, the main limitation is the training set, which beyond $i \approx 24$ rests mostly on the COSMOS-Web photometric redshifts. The code runs on CPUs and GPUs with identical results and inside RAIL. The \demp\ code will be made public on GitHub under the MIT license. The same method can therefore be applied to the larger and deeper training sets of the coming surveys.

\begin{acknowledgments}
I thank Yen-Ting Lin (ASIAA) and Keiichi Umetsu (ASIAA) for their suggestions and comments. The computations of this work were carried out on the SuMIRe Cluster at ASIAA. The SuMIRe Cluster is primarily funded by the Academia Sinica Investigator Award (AS-IA-107-M01 and AS-IA-112-M04), the National Science and Technology Council, Taiwan (NSTC 112-2112-M-001-027-MY3), and the Academia Sinica Career Development Award (AS-CDA-113-M01).

The Hyper Suprime-Cam (HSC) collaboration includes the astronomical communities of Japan and Taiwan, and Princeton University. The HSC instrumentation and software were developed by the National Astronomical Observatory of Japan (NAOJ), the Kavli Institute for the Physics and Mathematics of the Universe (Kavli IPMU), the University of Tokyo, the High Energy Accelerator Research Organization (KEK), the Academia Sinica Institute for Astronomy and Astrophysics in Taiwan (ASIAA), and Princeton University. Funding was contributed by the FIRST program from the Japanese Cabinet Office, the Ministry of Education, Culture, Sports, Science and Technology (MEXT), the Japan Society for the Promotion of Science (JSPS), Japan Science and Technology Agency (JST), the Toray Science Foundation, NAOJ, Kavli IPMU, KEK, ASIAA, and Princeton University.
This paper makes use of software developed for Vera C. Rubin Observatory. We thank the Rubin Observatory for making their code available as free software at \url{http://pipelines.lsst.io/}.
This paper is based on data collected at the Subaru Telescope and retrieved from the HSC data archive system, which is operated by the Subaru Telescope and Astronomy Data Center (ADC) at NAOJ. Data analysis was in part carried out with the cooperation of Center for Computational Astrophysics (CfCA), NAOJ. We are honored and grateful for the opportunity of observing the Universe from Maunakea, which has the cultural, historical and natural significance in Hawaii.

The Pan-STARRS1 Surveys (PS1) and the PS1 public science archive have been made possible through contributions by the Institute for Astronomy, the University of Hawaii, the Pan-STARRS Project Office, the Max Planck Society and its participating institutes, the Max Planck Institute for Astronomy, Heidelberg, and the Max Planck Institute for Extraterrestrial Physics, Garching, The Johns Hopkins University, Durham University, the University of Edinburgh, the Queen's University Belfast, the Harvard-Smithsonian Center for Astrophysics, the Las Cumbres Observatory Global Telescope Network Incorporated, the National Central University of Taiwan, the Space Telescope Science Institute, the National Aeronautics and Space Administration under grant No.\ NNX08AR22G issued through the Planetary Science Division of the NASA Science Mission Directorate, the National Science Foundation grant No.\ AST-1238877, the University of Maryland, Eotvos Lorand University (ELTE), the Los Alamos National Laboratory, and the Gordon and Betty Moore Foundation \citep{Chambers2016, Schlafly2012, Tonry2012, Magnier2013}.

This research used data obtained with the Dark Energy Spectroscopic Instrument (DESI). DESI construction and operations is managed by the Lawrence Berkeley National Laboratory. This material is based upon work supported by the U.S. Department of Energy, Office of Science, Office of High-Energy Physics, under Contract No.\ DE--AC02--05CH11231, and by the National Energy Research Scientific Computing Center, a DOE Office of Science User Facility under the same contract. Additional support for DESI was provided by the U.S. National Science Foundation (NSF), Division of Astronomical Sciences under Contract No.\ AST-0950945 to the NSF's National Optical-Infrared Astronomy Research Laboratory; the Science and Technology Facilities Council of the United Kingdom; the Gordon and Betty Moore Foundation; the Heising-Simons Foundation; the French Alternative Energies and Atomic Energy Commission (CEA); the National Council of Humanities, Science and Technology of Mexico (CONAHCYT); the Ministry of Science and Innovation of Spain (MICINN), and by the DESI Member Institutions: \url{www.desi.lbl.gov/collaborating-institutions}. The DESI collaboration is honored to be permitted to conduct scientific research on I'oligam Du'ag (Kitt Peak), a mountain with particular significance to the Tohono O'odham Nation. Any opinions, findings, and conclusions or recommendations expressed in this material are those of the author(s) and do not necessarily reflect the views of the U.S. National Science Foundation, the U.S. Department of Energy, or any of the listed funding agencies.

Funding for the Sloan Digital Sky Survey IV has been provided by the Alfred P. Sloan Foundation, the U.S. Department of Energy Office of Science, and the Participating Institutions. SDSS-IV acknowledges support and resources from the Center for High Performance Computing at the University of Utah. The SDSS website is \url{www.sdss4.org}. SDSS-IV is managed by the Astrophysical Research Consortium for the Participating Institutions of the SDSS Collaboration including the Brazilian Participation Group, the Carnegie Institution for Science, Carnegie Mellon University, Center for Astrophysics | Harvard \& Smithsonian, the Chilean Participation Group, the French Participation Group, Instituto de Astrof\'isica de Canarias, The Johns Hopkins University, Kavli Institute for the Physics and Mathematics of the Universe (IPMU) / University of Tokyo, the Korean Participation Group, Lawrence Berkeley National Laboratory, Leibniz Institut f\"ur Astrophysik Potsdam (AIP), Max-Planck-Institut f\"ur Astronomie (MPIA Heidelberg), Max-Planck-Institut f\"ur Astrophysik (MPA Garching), Max-Planck-Institut f\"ur Extraterrestrische Physik (MPE), National Astronomical Observatories of China, New Mexico State University, New York University, University of Notre Dame, Observat\'ario Nacional / MCTI, The Ohio State University, Pennsylvania State University, Shanghai Astronomical Observatory, United Kingdom Participation Group, Universidad Nacional Aut\'onoma de M\'exico, University of Arizona, University of Colorado Boulder, University of Oxford, University of Portsmouth, University of Utah, University of Virginia, University of Washington, University of Wisconsin, Vanderbilt University, and Yale University.
Funding for SDSS-III has been provided by the Alfred P. Sloan Foundation, the Participating Institutions, the National Science Foundation, and the U.S. Department of Energy Office of Science. The SDSS-III web site is \url{http://www.sdss3.org/}. SDSS-III is managed by the Astrophysical Research Consortium for the Participating Institutions of the SDSS-III Collaboration including the University of Arizona, the Brazilian Participation Group, Brookhaven National Laboratory, Carnegie Mellon University, University of Florida, the French Participation Group, the German Participation Group, Harvard University, the Instituto de Astrofisica de Canarias, the Michigan State/Notre Dame/JINA Participation Group, Johns Hopkins University, Lawrence Berkeley National Laboratory, Max Planck Institute for Astrophysics, Max Planck Institute for Extraterrestrial Physics, New Mexico State University, New York University, Ohio State University, Pennsylvania State University, University of Portsmouth, Princeton University, the Spanish Participation Group, University of Tokyo, University of Utah, Vanderbilt University, University of Virginia, University of Washington, and Yale University.
Funding for the SDSS and SDSS-II has been provided by the Alfred P. Sloan Foundation, the Participating Institutions, the National Science Foundation, the U.S. Department of Energy, the National Aeronautics and Space Administration, the Japanese Monbukagakusho, the Max Planck Society, and the Higher Education Funding Council for England. The SDSS Web Site is \url{http://www.sdss.org/}. The SDSS is managed by the Astrophysical Research Consortium for the Participating Institutions. The Participating Institutions are the American Museum of Natural History, Astrophysical Institute Potsdam, University of Basel, University of Cambridge, Case Western Reserve University, University of Chicago, Drexel University, Fermilab, the Institute for Advanced Study, the Japan Participation Group, Johns Hopkins University, the Joint Institute for Nuclear Astrophysics, the Kavli Institute for Particle Astrophysics and Cosmology, the Korean Scientist Group, the Chinese Academy of Sciences (LAMOST), Los Alamos National Laboratory, the Max-Planck-Institute for Astronomy (MPIA), the Max-Planck-Institute for Astrophysics (MPA), New Mexico State University, Ohio State University, University of Pittsburgh, University of Portsmouth, Princeton University, the United States Naval Observatory, and the University of Washington.

GAMA is a joint European-Australasian project based around a spectroscopic campaign using the Anglo-Australian Telescope. The GAMA input catalogue is based on data taken from the Sloan Digital Sky Survey and the UKIRT Infrared Deep Sky Survey. Complementary imaging of the GAMA regions was obtained by a number of independent survey programmes including GALEX MIS, VST KiDS, VISTA VIKING, WISE, Herschel-ATLAS, GMRT and ASKAP providing UV to radio coverage. GAMA is funded by the STFC (UK), the ARC (Australia), the AAO, and the participating institutions. The GAMA website is \url{https://www.gama-survey.org/}.

This work is based in part on observations made with the NASA/ESA/CSA James Webb Space Telescope, through the COSMOS-Web program (JWST-GO-1727; \citealt{Casey2023}) and the COSMOS2025 catalog \citep{Shuntov2025}, and on the COSMOS spectroscopic redshift compilation \citep{Khostovan2025}.
Based on data products from observations collected at the European Southern Observatory under ESO programs 070.A-9007 and 177.A-0837 (VVDS), 185.A-0791 (VUDS), 194.A-2003(E-T) (VANDELS) and 180.A-0776 (UDSz).
Funding for PRIMUS is provided by NSF (AST-0607701, AST-0908246, AST-0908442, AST-0908354) and NASA (Spitzer-1356708, 08-ADP08-0019, NNX09AC95G).
This research used computational resources from the Associa\c{c}\~ao Laborat\'orio Interinstitucional de e-Astronomia (LIneA) with the financial support of INCT do e-Universo (Process no.\ 465376/2014-2).
This research has made use of the VizieR catalogue access tool, CDS, Strasbourg, France \citep{Ochsenbein2000}.

Claude Opus 5.5 (Anthropic) was used in the preparation of this manuscript. It helped to draft and edit the text, to format the tables and the \LaTeX\ source, and to write the Python scripts that make the figures from the data described in this paper. I read, checked and edited all text, numbers, figures and references. Responsibility for the final manuscript lies entirely with the author.
\end{acknowledgments}

\facilities{Subaru (HSC), Mayall (DESI), Sloan, AAT, UKST (6dF), VLT:Melipal (VIMOS), Magellan:Baade (IMACS), JWST (NIRCam)}
\software{PyTorch \citep{Paszke2019}, NumPy \citep{Harris2020}, SciPy \citep{Virtanen2020}, Astropy (\citealt{Astropy2013}; \citealt{Astropy2018}; \citealt{Astropy2022}), scikit-learn \citep{Pedregosa2011}, Matplotlib \citep{Hunter2007}, RAIL \citep{RAIL}}

\bibliography{demp2026_pdr3_arxiv_20261126_v1}{}
\bibliographystyle{aasjournalv7}

\end{document}